\documentclass[preprint,12pt]{elsarticle}
\usepackage{graphicx}
\usepackage{epsfig}
\usepackage{booktabs}
\usepackage{threeparttable}
\usepackage{amssymb}
\usepackage{slashbox}
\usepackage{color}
\usepackage{enumerate}
\usepackage{subfigure}
\usepackage{url}
\usepackage{array}
\usepackage{pdflscape}
\usepackage{ulem}

\newcommand{\PreserveBackslash}[1]{\let\temp=\\#1\let\\=\temp}
\newcolumntype{C}[1]{>{\PreserveBackslash\centering}m{#1}}
\newcolumntype{R}[1]{>{\PreserveBackslash\raggedleft}p{#1}}
\newcolumntype{L}[1]{>{\PreserveBackslash\raggedright}p{#1}}
\makeatletter
\newcommand{\figcaption}{\def\@captype{figure}\caption}
\makeatother

\journal{Computational Biology and Chemistry}

\begin{document}
\begin{frontmatter}

\title{Human-chimpanzee alignment: Ortholog Exponentials and Paralog Power Laws}
\author{Kun Gao and Jonathan Miller}
\address{Physics and Biology Unit, Okinawa Institute of Science and Technology Graduate University, Okinawa, Japan}

\begin{abstract}

Genomic subsequences conserved between closely related species such as human and chimpanzee exhibit an exponential length distribution, in contrast to the algebraic length distribution observed for sequences shared between distantly related genomes. We find that the former exponential can be further decomposed into an exponential component primarily composed of orthologous sequences, and a truncated algebraic component primarily composed of paralogous sequences.

\end{abstract}


\end{frontmatter}

\section{Introduction}

Sequence conservation is defined by similar or identical nucleotide sequences within or among genomes at frequencies beyond those expected on neutral evolution. Within most neutral models of evolution, the probability that a sequence appears in two unrelated genomes decays exponentially with its length, so that sufficiently long sequences common to more than one genome are expected to derive from a common ancestor \cite{koonin}. Sequence duplication represents a primary mechanism through which new genetic material can arise \cite{lynch,ohno}. When identical sequences are observed within a single genome at levels exceeding those expected on an independent site model of evolution, sequence duplication is one candidate for their origin. Similarity among sequences beyond that expected within an independent site model, whether multiple occurrences within a single genome or simultaneous occurrence in multiple genomes, is known as ``sequence homology" and may indicate common ancestry \cite{homology}. 

Because sequence conservation and sequence duplication are often inferred from sequence length and identity, we believe that a systematic understanding of the latter two features may elucidate rules underlying sequence evolution and lead to more faithful models of neutral evolution.

The set of sequences shared within a genome or between two genomes may be summarised in its ``length distribution:" a histogram with length $L$ on the $x$-axis and number $\#(L)$ of shared sequences of length $L$ on the $y$-axis. These length distributions can exhibit distinctive characteristics that we aim to account for within some model of sequence evolution. Henceforth, we abbreviate ``length distribution" to `distribution," as all distributions referred to in this manuscript are histograms of the form indicated above.

\subsection*{Strong conservation among distantly related genomes: algebraic distribution with exponent $\approx -4$}

Distributions of sequences strongly conserved between a variety of distantly related genome pairs exhibit a heavy, approximately algebraic (power-law) tail \cite{SHM}. This power-law distribution is common but not universal; occurs not merely pairwise but also among multiple genomes; is robust over different measures of similarity; and has an exponent reported to be typically in the neighborhood of $-4$. Thus ``ultra-conserved" sequences exhibit this power law, but the same exponent also governs pairwise conserved sequences that are not necessarily shared by a third genome.

\subsection*{Sequence identity among closely related genomes: exponential distribution}

Sequences conserved between closely related species such as human and certain primates display an exponential distribution, rather than a power-law \cite{JMrep}. ``Closely related" is defined empirically as sufficiently recent branching from a common ancestor. 

\subsection*{``Ultra-duplication" within single genomes: algebraic distribution with exponent $-3$}

Study of exact duplications of all lengths in different genomes through whole-genome/whole-chromosome self-alignment revealed that duplicates with $100\%$ identity often -- but not always -- follow an approximately algebraic distribution with exponent in the neighbourhood of $-3$ \cite{GM}. Since it was originally observed (although with different exponent) in ultra-conserved sequences, in the context of duplicated sequences this algebraic feature was referred to as ``ultra-duplication," the prefix ``ultra" alluding solely to the long tail of the corresponding distribution, irrespective of its origin.

Massip and Arndt recently observed that together segmental duplication and point mutation can yield an algebraic distribution with exponent $-3$ \cite{stickbreak} (see also \cite{koroteev2013}). Customarily, segmental duplication is thought of as a neutral process, although selection may act subsequently. 

\begin{center} 
\makebox[\textwidth][c]{\includegraphics[width=1.3\textwidth]{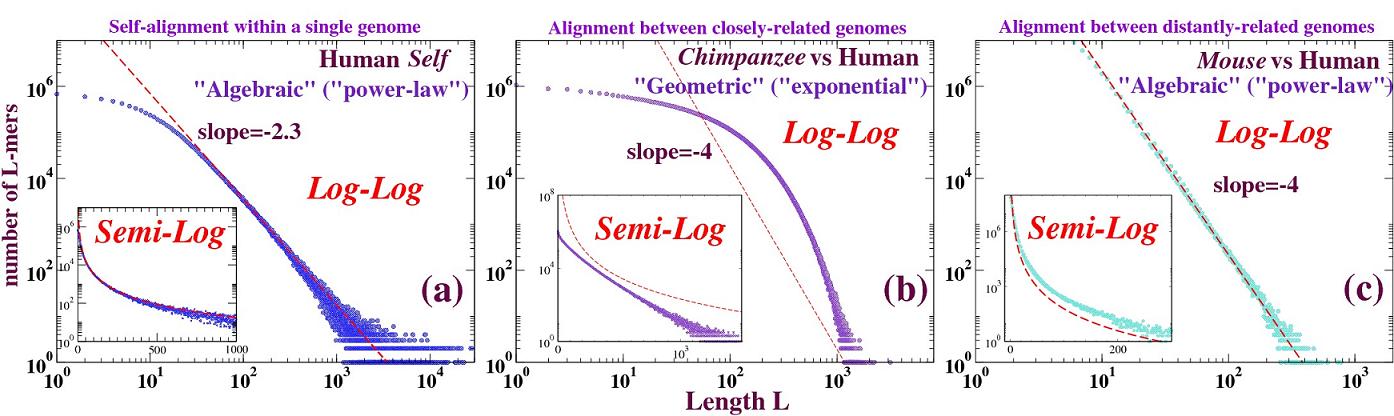}}
\figcaption{Distributions of identically conserved or duplicated sequences in (a) human chromosome $1$ self-alignment; (b) human chromosome $1$ -- chimpanzee chromosome $1$ alignment and (c) human chromosome $1$ -- mouse chromosome $1$ alignment. Distributions shown are generated by repeat-masked whole-chromosome LASTZ \cite{lastz} net alignments obtained directly from UCSC Genome Bioinformatics. Log-log plots enclose semi-log insets.}
\label{hmp}
\end{center}

Figure \ref{hmp} recapitulates the three cases mentioned above. With increasing evolutionary distance between species, distributions of identical sequences obtained from LASTZ net alignment cross over from algebraic (with exponent $-3$) to exponential, and then again to algebraic (with exponent in the neighborhood of $-4$). These crossovers are further elucidated below. All alignments described in this manuscript were performed with LASTZ (see \textbf{Materials and Methods}); henceforth -- with the exception of the ``\textbf{Materials and Methods}" section -- we refer to ``alignment" and for the most part we omit the qualifier ``LASTZ," which is tacitly implied unless otherwise indicated explicitly.  \\

In the following, we apply whole-genome/whole-chromosome alignment between human and chimpanzee to investigate the origin of the exponential distribution and disentangle it from the algebraic distribution. For closely related species, quantitative relationships emerge between orthologous sequences and the exponential distribution, and between paralogous sequences and the algebraic distribution.

\section{Materials and Methods}

\subsection{Pairwise alignment of genome sequences}

\subsubsection*{Software}

We compare genomic sequences with the LASTZ pairwise alignment tool \cite{lastz}. LASTZ alignment comprises several stages of which we rely mainly on two: \textit{raw} alignment and \textit{net} alignment. Raw alignment is the immediate product of LASTZ and may include multiple and positionally overlapping matches for each aligned sequence. A subsequent net alignment removes positional overlaps among matched sequences, chains them, and discards all but the highest-scoring chains, yielding a single match for each position in the genome. One function of net alignment is to extract homologous elements from the raw alignment \cite{Kent}.

LASTZ is obtained from \url{http://www.bx.psu.edu/miller_lab/}; we use LASTZ default options for raw alignment. The UCSC Genome Browser (\url{http://hgdownload.cse.ucsc.edu/admin/jksrc.zip}) provides additional tools (\textit{axtChain}, \textit{chainNet} and \textit{netToAxt}) for producing the net alignment. Standard procedures that we follow for LASTZ alignment (both raw and net) are described at: \url{http://genomewiki.cse.ucsc.edu/index.php/Whole_genome_alignment_howto}.

\subsubsection*{Genome sequences}

Soft repeat-masked (\url{http://repeatmasker.org}) genome sequences are obtained as fasta files from the Ensembl FTP Server (e.g. $hg19$ as version 74; \url{ftp://ftp.ensembl.org/pub/release-74/fasta/homo_sapiens/dna/Homo_sapiens.GRCh37.74.dna.chromosome.1.fa.gz}). We use for the most part the $hg19$ and $panTro4$ assemblies for human and chimpanzee respectively. For most of our calculations, we study the human chromosome $1$ -- chimpanzee chromosome $1$ LASTZ raw alignment.

Other primate genomes are obtained from Ensembl, and human chromosome $1$ is aligned to its respective ``orthologous" counterpart from each primate -- the primate chromosome that shares with human chromosome $1$ the most orthologous genes as identified in Ensembl Biomart (\url{http://www.ensembl.org/biomart/martview}), yielding gorilla chromosome $1$, orangutan chromosome $1$ (reverse strand), macaca chromosome $1$, and marmoset chromosome $7$ as orthologous to human chromosome $1$.

Mouse ($mm9$) chromosome $1$ is downloaded from Ensembl and aligned to human chromosome $1$. Mouse chromosome $1$ carries a plurality (close to $1/4$) of orthologous elements shared between human chromosome $1$ and the mouse whole genome. The Venter genome \cite{venter} is obtained from UCSC Genome Bioinformatics (\url{http://hgdownload.cse.ucsc.edu/goldenPath/venter1/bigZips/venter1.2bit}) and aligned to $hg19$.

\subsubsection*{Repeat-masking}

Repetitive sequence elements may constitute close to half the genome; unless these repeats are explicitly identified, most if not all large-scale alignment methods may fail to complete on eukaryotic genomes. A common practice is to identify these elements prior to alignment with a software tool such as Repeatmasker (\url{http://repeatmasker.org}) that demarcates repetitive sequence in lower-case letters (``soft [repeat-]masking") in contrast to the upper-case letters that designate non-repetitive sequence.

\textit{Unless otherwise indicated, all LASTZ alignments represented here are performed on soft repeat-masked genome sequences.}

When aligning sequences, LASTZ excludes soft-masked bases from its ``seed" stage but reintroduces them in later stages, when alignments of unmasked sequence can be extended into soft-masked regions (see webpage: \url{http://www.bx.psu.edu/miller_lab/dist/README.lastz-1.02.00/README.lastz-1.02.00a.html#overview}), so that LASTZ can in principle align certain masked sequences. For example, just over $50\%$ of each of human ($hg19$; Ensembl GRCh$37.74$) chromosome $1$ and chimpanzee (Ensembl CHIMP$2.1.4.74$) chromosome $1$ is repeat-masked. Nevertheless, the LASTZ raw alignment between these two soft repeat-masked chromosomes covers $94\%$ of the human chromosome, and $97\%$ of the chimpanzee chromosome; $92\%$ of the masked bases in human chromosome $1$ and $96\%$ of the masked bases in chimpanzee chromosome $1$ are aligned by LASTZ.

\subsection{Parsing the alignment}

\subsubsection*{Distribution of contiguous matched runs (CMRs)}

Following \cite{GM}, for a given pairwise alignment we study continuous (uninterrupted) matching runs of bases (CMRs), wherein a contiguous series of matching nucleotides is terminated at mismatches or indels. Unless otherwise indicated, all CMRs discussed here represent exact matches that we refer to interchangeably with these two terms.

A histogram $\#(L)$ (or (length) distribution) describes the number of CMRs of a given length $L$. Pairwise alignment of genomes yields conserved or duplicated sequences within or between genomes; we expect that distributions of these conserved or duplicated sequences reflect certain features of genome evolution.

\subsubsection*{Forward alignment and reverse alignment}

DNA is composed of complementary strands so that for two DNA sequences pairwise alignment can be implemented in either of two relative orientations, ``forward" or ``reverse." Matches to the reverse orientation are thought to arise by inversion or inverted duplication/transposition \cite{inversion}. We perform both forward and reverse alignments; however, we subsequently combine their products before further calculation except where it is informative to keep them separate.

\subsubsection*{Dot plot}

A two-dimensional similarity matrix between two sequences is displayed as a dot plot \cite{dp}, in which one sequence of an aligned pair lies along the horizontal and the other along the vertical axis. Dot plots are commonly used to visualise sequence similarity and to display homologous matches between genomes. ``Syntenic dot plots" exhibit synteny (see webpage: \url{http://genomevolution.org/wiki/index.php/Syntenic_dotplots}). In this paper, we apply them to display orthologous sequences between human and chimpanzee genomes.

In some of our dot plots, prominent horizontal or vertical white bands appear that correspond to sequence that has not yet been reliably determined and is therefore represented by ``N" in the assemblies from Ensembl; such bases are excluded from alignments \cite{lastz}.

\section{Results}

\subsection{Alignments of orthologous regions of genomes yield exponential distributions}

An alignment of a numbered human chromosome to the chromosomes of chimpanzee yields the exponential distribution of CMRs shown in figure \ref{hmp} for the correspondingly numbered chimpanzee chromosome, but not for other chimpanzee chromosomes.  In the latest releases of the chimpanzee genome assembly, chimpanzee chromosomes have been renumbered to reflect common ancestry with the corresponding human chromosomes \cite{OC}, so that chromosomes sharing the same number can be thought of as ``orthologous chromosomes." In figure \ref{panall}, human chromosome $1$ is separately aligned against each chimpanzee chromosome. With the exception of the alignment of human chromosome $1$ with chimpanzee chromosome $1$, all the alignments (raw and net) between human chromosome $1$ and chimpanzee chromosomes yield approximately algebraic distributions. The dot plots corresponding to these alignments can be found in figure \ref{panalldp}.

\begin{center} 
\makebox[\textwidth][c]{\includegraphics[width=\textwidth]{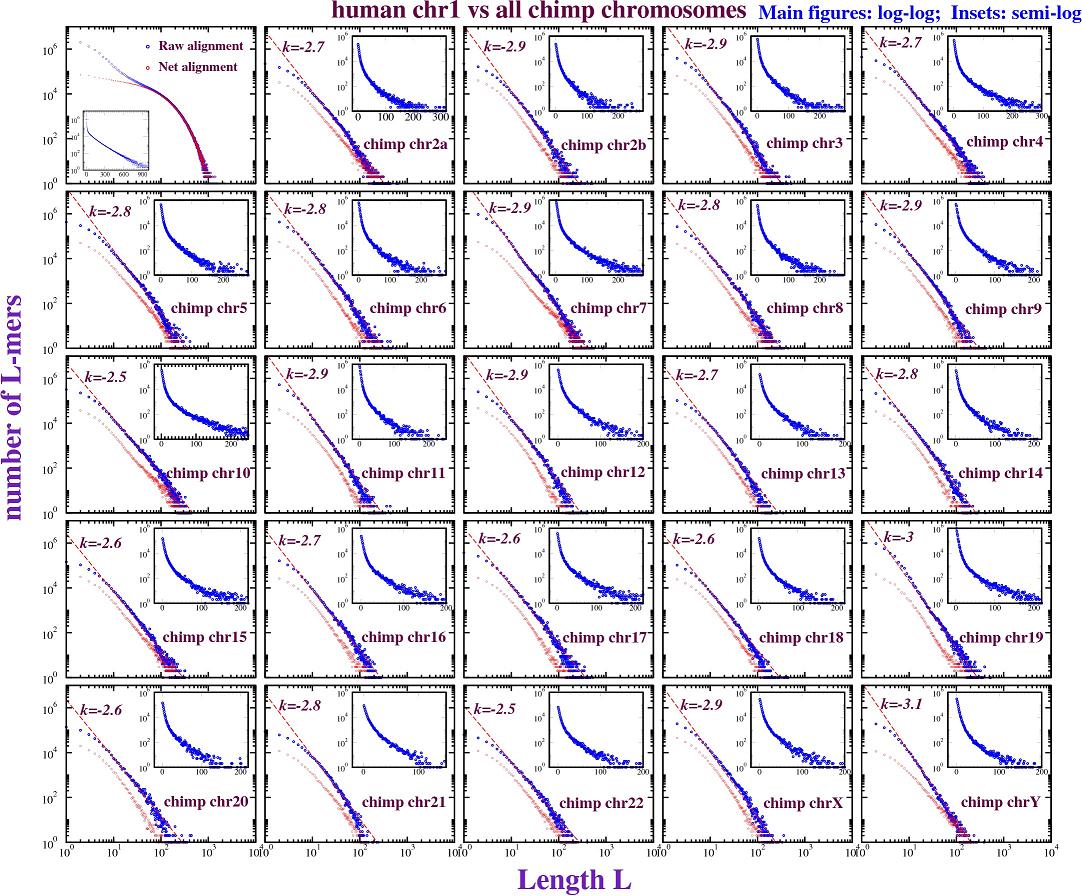}}
\figcaption{Distributions of exact matches (CMRs) in raw (blue) and net (red) alignments of human chromosome $1$ against all chimpanzee chromosomes. Main figures show log-log plots; insets semi-log plots for the same data. For purposes of comparison, lines with slope $k$ on the log-log scale as indicated have been drawn into each figure.}
\label{panall}
\end{center}

Although human chromosome $1$ -- chimpanzee chromosome $1$ alignment exhibits an exponential distribution overall, figure \ref{hplocal} suggests that this exponential is composed solely or primarily of CMRs between orthologous regions of these two chromosomes. Figure \ref{hplocal} illustrates alignments of orthologous versus heterologous sequences in human chromosome $1$ and chimpanzee chromosome $1$. $H\_frag1$, $H\_frag2$, $C\_frag1$ and $C\_frag2$ are fragments taken respectively from the first and last thirds of these two chromosomes. For these four fragments, figure \ref{orthmap} shows an orthology map and figure \ref{synmap} a syntenic dot plot. As can be seen in figure \ref{hplocal}, alignments between homologous (heterologous) fragments exhibit exponential (algebraic) distributions.

\begin{center} 
\includegraphics[width=0.75\textwidth]{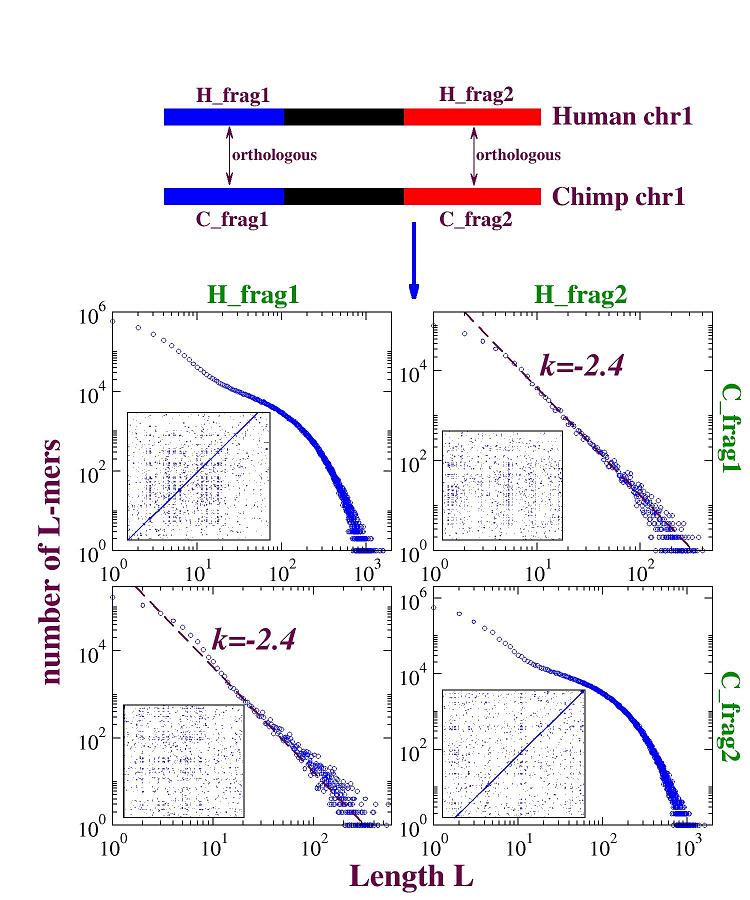}
\figcaption{Distributions of exact matches in raw alignments between different fragments of human chromosome $1$ and chimpanzee chromosome $1$. We extract the first and last thirds of each these two chromosomes, and align all four resulting fragment pairs. These figures indicate that even within a single chromosome, the exponential distribution correlates with orthology: only alignments between orthologous fragments show exponential distributions.} 
\label{hplocal}
\end{center}

\begin{center} 
\includegraphics[width=0.5\textwidth]{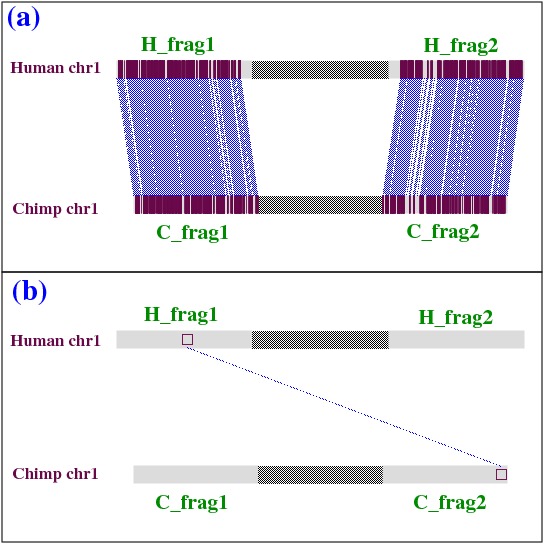}
\figcaption{Orthology map among different fragments of human chromosome $1$ and chimpanzee chromosome $1$. Horizontal dark grey bars (largely obscured in (a) by maroon vertical bars) show human chromosome $1$ and chimpanzee chromosome $1$; light grey bars indicate fragments $H\_frag1$, $H\_frag2$, $C\_frag1$ and $C\_frag2$ defined in figure \ref{hplocal}. Maroon vertical bars indicate the locations genes orthologous between human chromosome $1$ and chimpanzee chromosome $1$; and blue dotted lines connect these orthologous gene pairs.}
\label{orthmap}
\end{center}

We describe further evidence below that the exponential distribution observed in the human-chimpanzee alignment correlates with sequence orthology. 

\subsection{Separating the exponential from the algebraic}

Although alignment of orthologous human and chimpanzee chromosomes yields an exponential distribution overall, distributions of aligned subfragments of these genomes are not necessarily exponential. In figures \ref{panall} and \ref{hplocal} it is seen that whole-chromosome alignments between human and chimpanzee contain both exponential and power-law components.

In this subsection, we illustrate how the human chromosome $1$ -- chimpanzee chromosome $1$ alignment naturally decomposes into algebraic and exponential subsets. Based on the observations in figure \ref{panall} and figure \ref{hplocal}, we hypothesise that \\

\textit{\textbf{to a first approximation, the exponential and (approximately) power-law components correspond to orthologous sequence and paralogous sequence, respectively.}}\\

To perform this decomposition we develop several methods, each of which is related to this hypothesis in a slightly different way. With the exception of a ``local" method based on ``nested" and ``non-nested" matches that is parameter-free, they involve parameter choices and sometimes further manipulations whose justification is not always readily apparent. Nevertheless, it turns out that these methods yield very similar outcomes. 

It may be worth remarking that a length distribution alone contains no information about location in a genome, so that it is impossible to partition an alignment into exponential and algebraic components solely on the basis of aligned fragment or CMR lengths; nevertheless, the content of the previous subsection suggests that a partition can be extracted from the dot plot.

\subsubsection{``Geometrical" method: Separating on-diagonal elements from off-diagonal elements}
\label{ms1}

As evident from figure \ref{panall}, figure \ref{hplocal} and figure \ref{panalldp}, alignments between human and chimpanzee genomes with an exponential distribution exhibit dense accumulations of sequences within the dot plot. For closely related species like human and chimpanzee, it is well known that one such high density zone ordinarily forms a band near the diagonal of the dot plot that we refer to as the ``diagonal band." We will see that the diagonal band is a major contributor to the exponential distribution.

Figure \ref{onoff} shows the distributions of exact matches from the diagonal band and from its complement within the dot plot of the human chromosome $1$ -- chimpanzee chromosome $1$ raw alignment. We crudely take into account the length difference between human chromosome $1$ and chimpanzee chromosome $1$, on the order of $\delta_C1 = 10^7$ bases, by defining a region around the diagonal of width $\delta_C1$, so that sequences offset by as many as $\delta_C1$ bases are to be thought of -- in this approximation --  as comprising the diagonal band. The exponential distributions of CMRs in this diagonal band and the algebraic distributions of the off-diagonal CMRs are evident in the right-most panels of figure \ref{onoff}.

\begin{center} 
\makebox[\textwidth][c]{\includegraphics[width=1.2\textwidth]{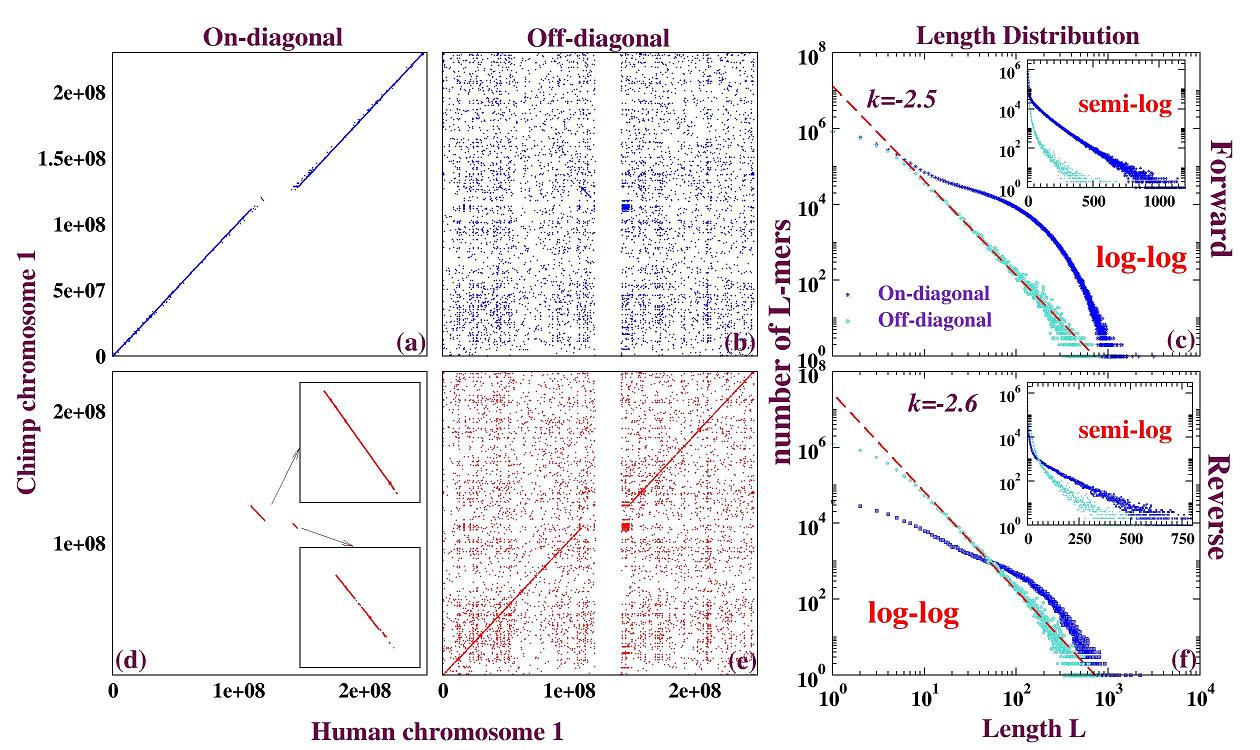}}
\figcaption{Dot plots and distributions of exact matches on diagonal and off diagonal in human chromosome $1$ -- chimpanzee chromosome $1$ raw alignment. The diagonal band width is chosen as $\delta_C1 = 10^7$ bases (see text). For the reverse alignment, we excise not the whole diagonal band, but rather only two fragments (see the insets in the lower-left panel) that in the dot plot correspond to large inversions. The exponential distribution of CMRs in this diagonal band and the algebraic distribution of off-diagonal CMRs is evident in the right-most panels.}
\label{onoff}
\end{center}

In contrast to the other methods described here, in this subsection we treat the forward and reverse alignment separately. As we have discussed above, the orthologous sequences between human chromosome $1$ and chimpanzee chromosome $1$ concentrate primarily in the forward strands; we extract the entire diagonal band from the forward alignment. In the reverse alignment, two large and distinct inversions appear on the dot plot (see insets in lower-left panel of Figure \ref{onoff}); by extracting these inversions we find empirically that we can neatly separate exponential from power-law in the reverse alignment. This can be understood if these large inversions are recent events, so that in contrast to the rest of these two chromosomes, the orthologous orientation is reversed.

\subsubsection{``Genetic clock" method: Separating high-similarity alignment blocks from low-similarity alignment blocks}
\label{ms2}

The raw alignment is composed of a set of alignment blocks, each representing a local alignment whose score is higher than a pre-established threshold. One way to characterise similarity in an alignment block is to compute the number of mismatches it contains, yielding a Hamming distance. The ratio of Hamming distance to sequence length then represents a (time-integrated) rate of variation per base. In the absence of selection, and under customary idealisations, this ratio reflects the time elapsed subsequent to divergence, constituting a crude ``genetic clock" \cite{geneclock}. According to the definition of ortholog and paralog (see section \textit{\ref{oplg}}), paralogs diverge before orthologs and should exhibit greater ratios of Hamming distance to sequence length than orthologs. Figure \ref{DMR} left panel shows Hamming distance versus alignment block length for all alignment blocks in the human chromosome $1$ -- chimpanzee chromosome $1$ raw alignment; each dot corresponds to an alignment block. Evidently, these alignment blocks comprise two major branches; alignment blocks in the lower-right branch exhibit lower rates of variation (greater similarity) than those in the upper-left branch.

Figure \ref{DMR} left panel shows a natural partition of alignment blocks; a line with slope in the neighborhood of $0.08$ from the origin is sufficient to elucidate a partition into an upper-left branch and a lower-right branch (leftmost panel). Middle panels in figure \ref{DMR} show respective dot plots for the alignment blocks in the upper-left branch and lower-right branch, and the rightmost panel the (approximately) algebraic distribution of the upper-left branch and the exponential distribution of the lower-right branch.

In this ``genetic clock" method, we treat the forward and reverse alignments identically, thus in figure \ref{DMR}, we display the combined product of forward and reverse alignment; they are exhibited separately in figure \ref{DMRFI}.

\begin{center} 
\makebox[\textwidth][c]{\includegraphics[width=1.3\textwidth]{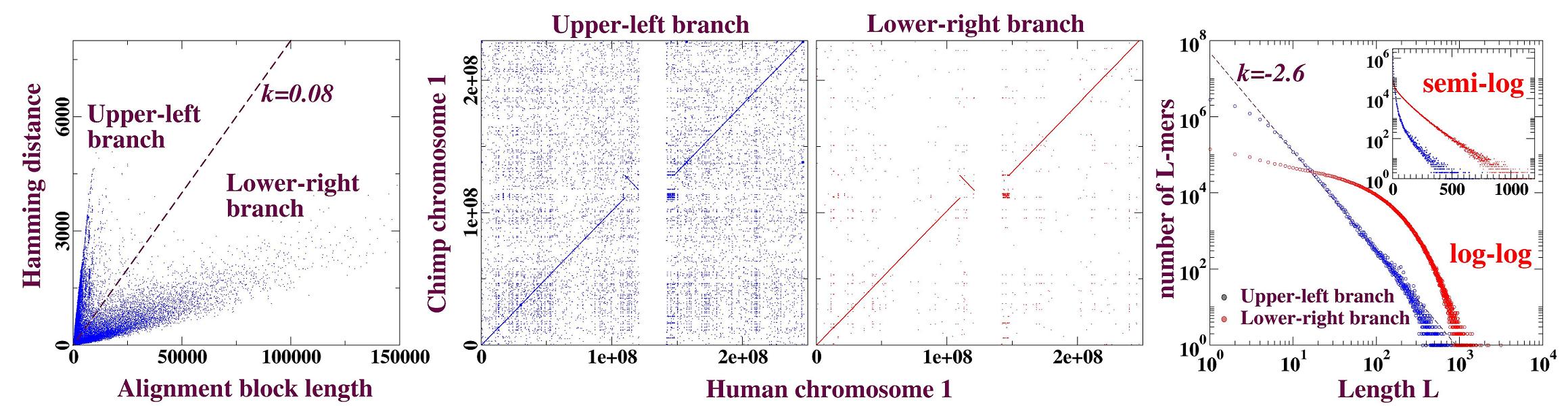}}
\figcaption{Dot plots and distributions of exact matches depend on the accumulated variation (see text) within the alignment blocks from which they are derived. From raw alignment between human chromosome $1$ and chimpanzee chromosome $1$, the left panel shows the Hamming distance (number of mismatches and indels) as a function of alignment block length; each dot corresponds to a distinct alignment block. The dashed lines crudely partition alignment blocks into an upper-left branch and a lower-right branch. Dot plots of exact matches for alignment blocks within each branch are shown in the middle panels, and their respective distributions in the right panel, exhibiting decomposition into (approximately) algebraic and exponential components.}
\label{DMR}
\end{center}

\subsubsection{``Global" method: Extracting the net alignment from the raw alignment}
\label{ms3}

LASTZ alignment is performed in stages, with ``raw" alignment the immediate product. Raw alignment contains all matches between sequences whose alignment scores exceed a predetermined threshold. Aligned fragments often overlap within the raw alignment; one location in the target sequence can match multiple locations in the query sequence, and vice versa. Net alignment scans the target sequence and selects from the aligned fragments in each region the pair with the highest alignment score, discarding all pairs with lower scores, eliminating overlaps and returning a unique optimal chain of aligned fragments \cite{Kent}.

Since the exponential distribution of CMRs in human-chimpanzee alignment comprises primarily of high-similarity sequence pairs, one would expect the net alignment to extract such pairs from the raw alignment. We define a ``raw minus net" (RMN) alignment as the residual of the raw alignment once all fragments also in the net alignment have been removed. Thus the net alignment and the RMN alignment represent complementary subsets of the raw alignment.

Kent et al. designed the net alignment to align orthologous sequences \cite{Kent}, so it is not surprising that the LASTZ net alignment between human chromosome $1$ and chimpanzee chromosome $1$ consists primarily of exponential components. 

Figure \ref{RMN} exhibits dot plots and distributions of exact matches for raw, net and RMN alignments between human chromosome $1$ and chimpanzee chromosome $1$; and it can be seen there that the net alignment extracts an exponential component from the raw alignment; the RMN alignment distills an (approximately) algebraic component. Figure \ref{RMNFI} exhibits these plots and distributions separately for forward alignment and for reverse alignment.

\begin{center} 
\makebox[\textwidth][c]{\includegraphics[width=1.3\textwidth]{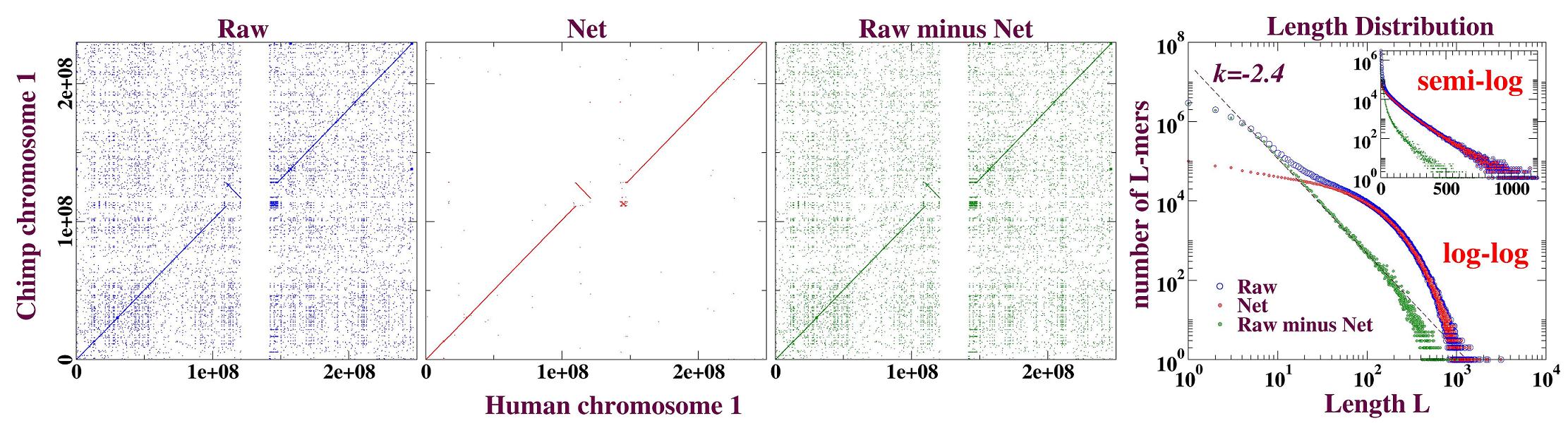}}
\figcaption{Dot plots and distributions of exact matches for the raw, net and raw minus net alignments between human chromosome $1$ and chimpanzee chromosome $1$. The net alignment extracts an exponential component from the raw alignment; the RMN alignment distills an (approximately) algebraic component.}
\label{RMN}
\end{center}

\subsubsection{``Local" method: Separating non-nested-CMRs from nested-CMRs}
\label{ms4}

We define ``nested-CMRs" and ``non-nested-CMRs" as two complementary subsets of the CMRs within an alignment: a CMR is said to be ``nested" if it is a subsequence of another CMR. In more detail,\\

\noindent \textbf{\textit{Definition 1:}} If \textbf{\textit{seq:}} \mbox{\boldmath$[i_1,i_2]$} denotes a sequence that starts at location \mbox{\boldmath$i_1$} and ends at location \mbox{\boldmath$i_2$} in a genome (here \mbox{\boldmath$i_2\ge i_1$} are coordinates in the genome, both relative to the plus strand), then for two sequences extracted from a same genome, \textbf{\textit{seqA:}} \mbox{\boldmath$[x_1,x_2]$} and \textbf{\textit{seqB:}} \mbox{\boldmath$[y_1,y_2]$}, we say ``\textbf{\textit{seqA}} is nested in \textbf{\textit{seqB}}" if both these conditions are satisfied:

\begin{enumerate}
\item \mbox{\boldmath$y_2-y_1 \geq x_2-x_1$};
\item \mbox{\boldmath$y_1 \leq x_1$};
\item \mbox{\boldmath$y_2 \geq x_2$};
\end{enumerate}

\noindent \textbf{\textit{Definition 2:}} Given two different CMRs within an alignment, when the query or target sequence of one CMR is nested in the corresponding query or target sequence of the other, we say the first CMR is nested in the second CMR, and the overlap between these two CMRs is called a ``nested overlap."\\

\noindent \textbf{\textit{Definition 3:}} A CMR that is nested in another CMR is called a \textbf{\textit{nested-CMR}}; otherwise it is a \textbf{\textit{non-nested-CMR}}.\\

These definitions of nested and non-nested CMR apply to any alignment, including -- but not limited to -- LASTZ raw \cite{lastz} and LASTZ net \cite{Kent} \footnote{However, please note that here our definitions of nested and non-nested CMRs are different from those of nested and non-nested local maxmers by E Taillefer and J Miller in \cite{TM2014}.}. Below, we apply these definitions to LASTZ raw alignment, and study the distributions exhibited by nested and non-nested CMRs.

\begin{center} 
\makebox[\textwidth][c]{\includegraphics[width=1.1\textwidth]{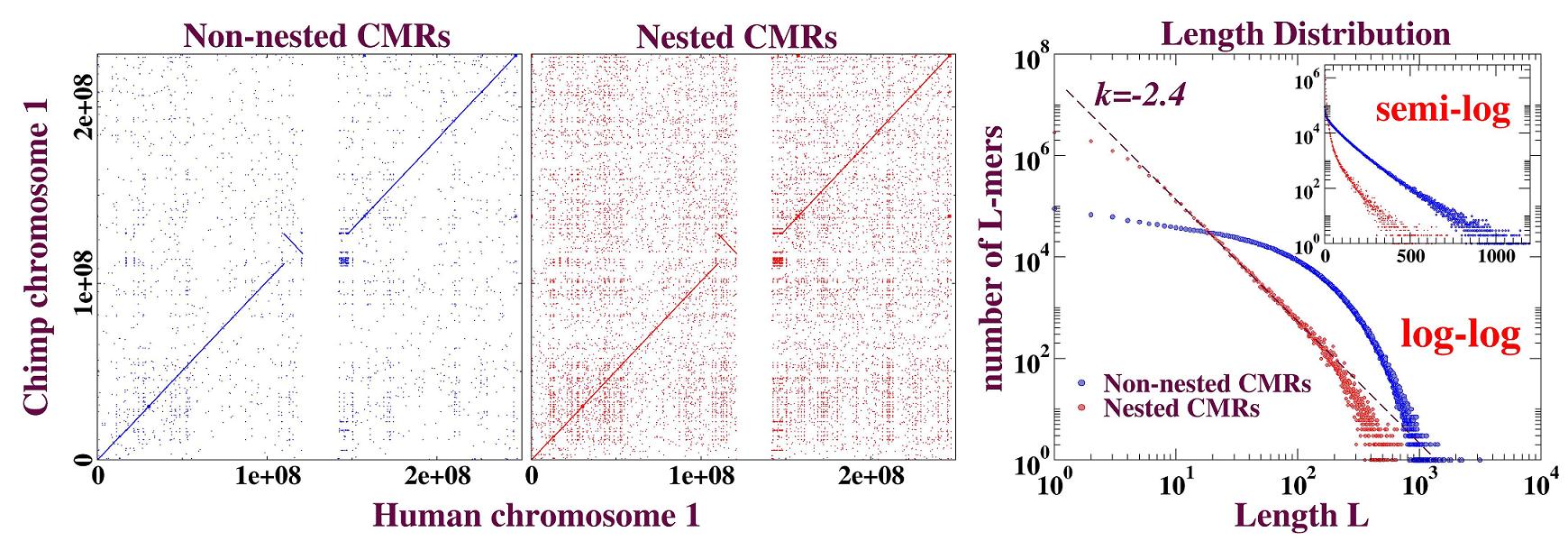}}
\figcaption{Dot plots and distributions of exact matches for  nested-CMRs and non-nested-CMRs in human chromosome $1$ -- chimpanzee chromosome $1$ raw alignment.}
\label{neov}
\end{center}

Figure \ref{neov} exhibits dot plots and distributions for the nested-CMRs and non-nested-CMRs in human chromosome $1$ -- chimpanzee chromosome $1$ raw alignment (for forward and reverse alignments alone see figure \ref{neovfi}). The (approximately) algebraic character of nested-CMRs versus the exponential character of non-nested-CMRs is evident. This outcome is plausible if one recalls that orthologs tend to be more similar to one another than are paralogs (see section \textit{\ref{oplg}}), so that subsequences of paralogs are likely to be nested within subsequences of orthologs. This method requires no chaining of alignment blocks, and is further distinguished from netting because it is parameter-free.

\subsubsection{Different methods are consistent with one another}

Aside from their common reliance on the raw alignment, these four methods (\textit{\ref{ms1}} -- \textit{\ref{ms4}}) are independent of one another; however, the distributions of the corresponding subsets extracted by each of these four methods are largely similar. Differences are only apparent in the dot plots. For example, to obtain the exponentially distributed subset, method \textit{\ref{ms1}} extracts the entire diagonal band, discarding all off-diagonal elements. In contrast, the other methods all retain some on-diagonal and some off-diagonal elements.

Figure \ref{cmro} schematically displays the consistency of these methods, indicating that exponential subsets extracted by different methods consist overwhelmingly of shared CMRs; in particular our ``global" and ``local" methods share close to $95\sim98\%$ of the CMRs (figure \ref{cmro} left panel). 

\begin{center} 
\includegraphics[width=0.47\textwidth]{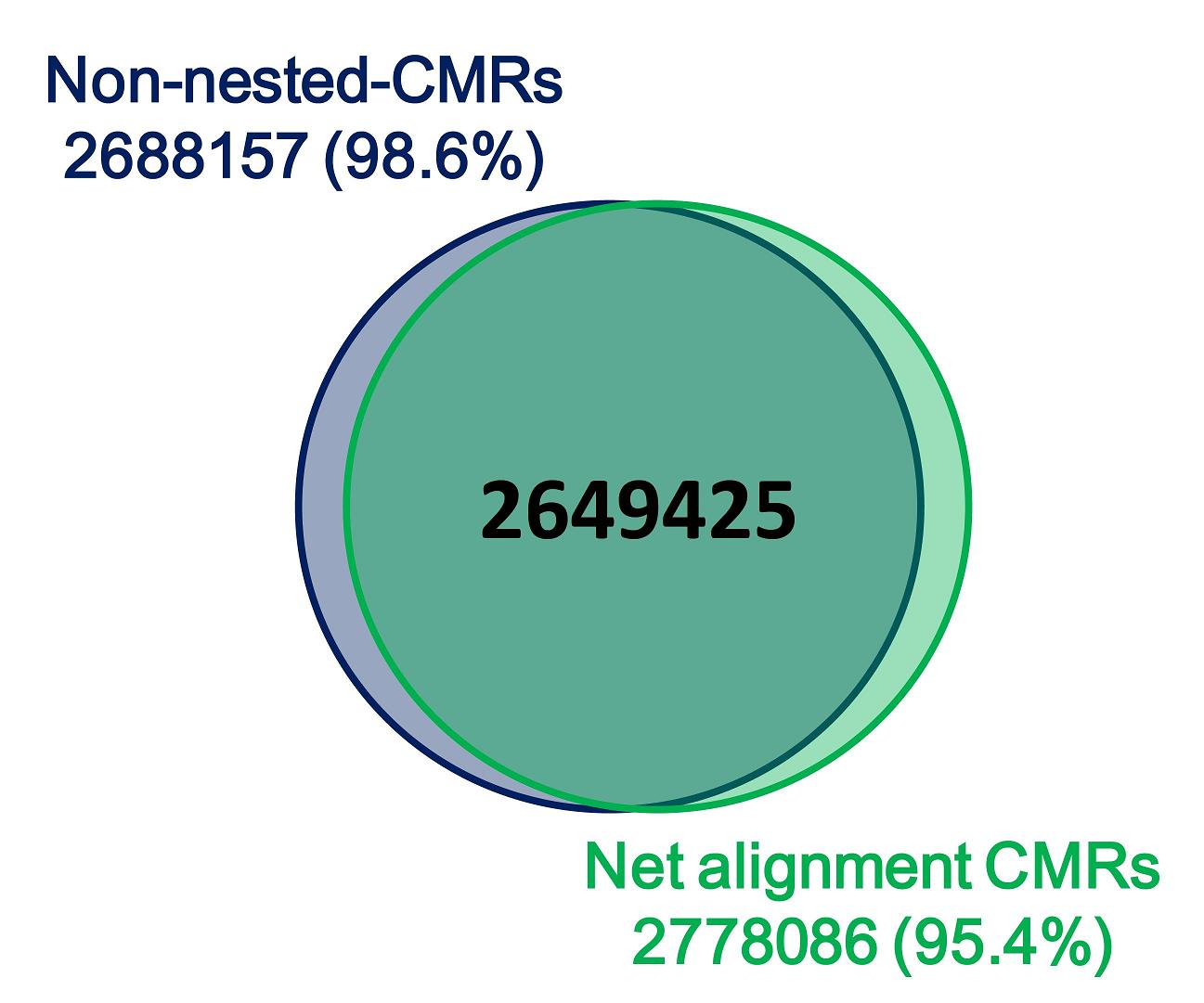}
\includegraphics[width=0.47\textwidth]{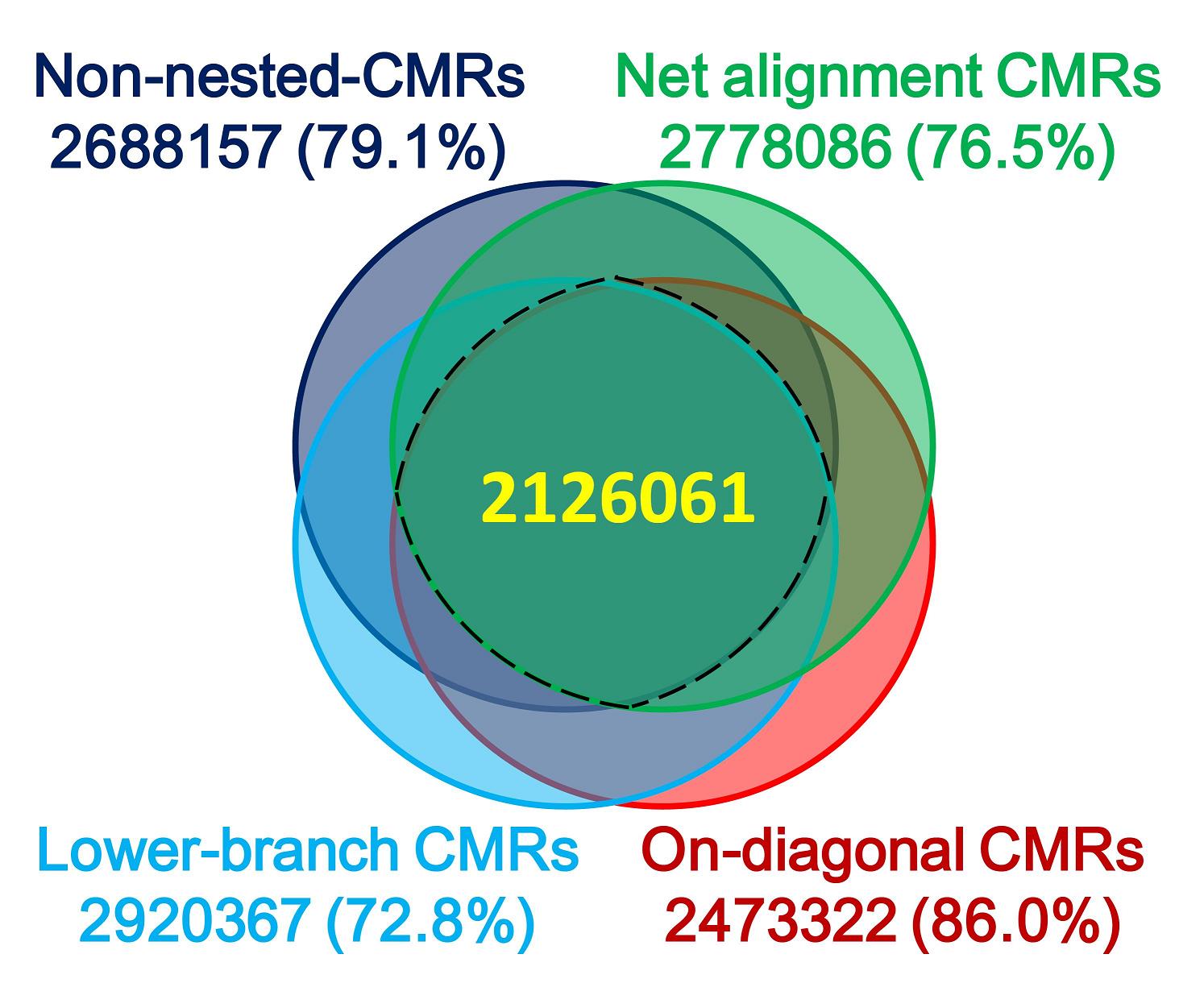}   
\figcaption{Schematic illustrations of consistency among different methods \textit{\ref{ms1}} -- \textit{\ref{ms4}} described in the text. Circles in the figure indicate the exponential subsets extracted from human chromosome $1$ -- chimpanzee chromosome $1$ raw alignment. Numerals in the figure show the number of CMRs in different subsets and percentages in the brackets show the proportions of shared CMRs.}
\label{cmro}
\end{center}

As evident in the right panel of figure \ref{cmro}, the set of CMRs common to all four methods contains at least $70\%$ of the CMRs obtained by each method alone. Although each of these four methods yields some CMRs that are not obtained by any of the other methods, the proportions of such CMRs are small: $1.6\%$ of the net alignment CMRs, $0.4\%$ of the non-nested CMRs, $6.2\%$ of the low-branch CMRs and $7.8\%$ of the on-diagonal CMRs.

\subsection{Random uncorrelated point mutation (RUPM) model}

To account qualitatively for the ortholog contribution to the exponential distribution, we apply a random uncorrelated point mutation (RUPM) model. As a simple model of neutral evolution, a RUPM model consists of site-independent point mutations (here, single-base substitutions) only, where the rate of these mutations is homogeneous across the genome. 

As two identical copies of a common ancestor genome evolve independently under a neutral RUPM model, CMR lengths follow an exponential distribution. For sufficiently short times, long CMRs can be assigned to corresponding positions within the two genomes, and lie on the diagonal; long segmental duplications present in the common ancestor remain well conserved. Matches among these segmental duplications in different locations of the genomes yield a distribution similar to that of the common ancestor: any differences can only be accounted for by random, uncorrelated point mutation.

\subsubsection{A synthetic alignment under the RUPM model}
\label{synsim}

We perform a numerical simulation of neutral evolution under the RUPM model. Human chromosome $1$ was selected as a common ancestor sequence containing algebraically distributed segmental duplications. Starting from two identical copies of the ancestral genome, random uncorrelated point mutations are introduced independently. We apply $0.5\%$ mutations per base and generate a raw alignment between the descendent genomes. The distributions of exact matches are displayed in figure \ref{synthe}.

\begin{center} 
\includegraphics[width=0.7\textwidth]{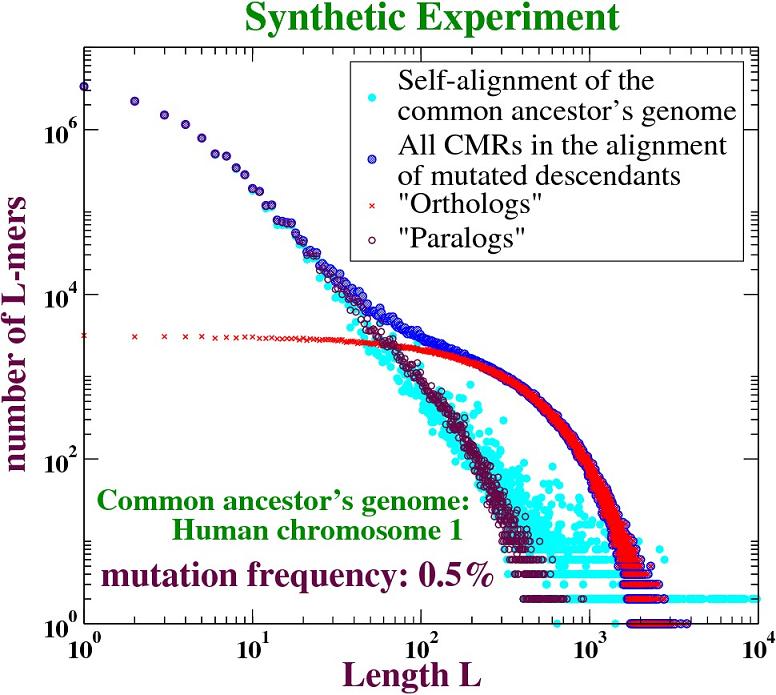}
\figcaption{Distributions of exact matches in raw alignment between two ``synthetic" descendants of a common ancestor genome. Introducing random uncorrelated point mutations with frequencies $0.5\%$ into two copies of an ancestral genome consisting of human chromosome $1$ generates two synthetic descendent genomes. In the figure, solid cyan circles indicate self-alignment of the original (un-mutated) sequence; solid blue circles all the CMRs in the alignment between the mutated sequences; red crosses the ``orthologs;" open maroon circles the ``paralogs." ``Orthologs" correspond to matches that share common locations between the two descendent genomes' ``paralogs" to matches with a different location in each of the two descendent genomes.}
\label{synthe}
\end{center}

Under the RUPM model, we identify matches between sequences having identical coordinates within the respective mutated sequences as ``orthologs" and all other matches as ``paralogs." In figure \ref{synthe} these orthologs exhibit an exponential distribution, whereas paralogs exhibit an (approximately) algebraic distribution that resembles the algebraic distribution of the self-alignment of the original (un-mutated) sequence, but falls a little short in the tail.

For comparison, a parallel simulation on a random sequence is performed; see \ref{ctrlsim}.

\subsubsection{Separating orthologs from paralogs with different methods in the synthetic alignment}

\begin{center} 
\includegraphics[width=0.7\textwidth]{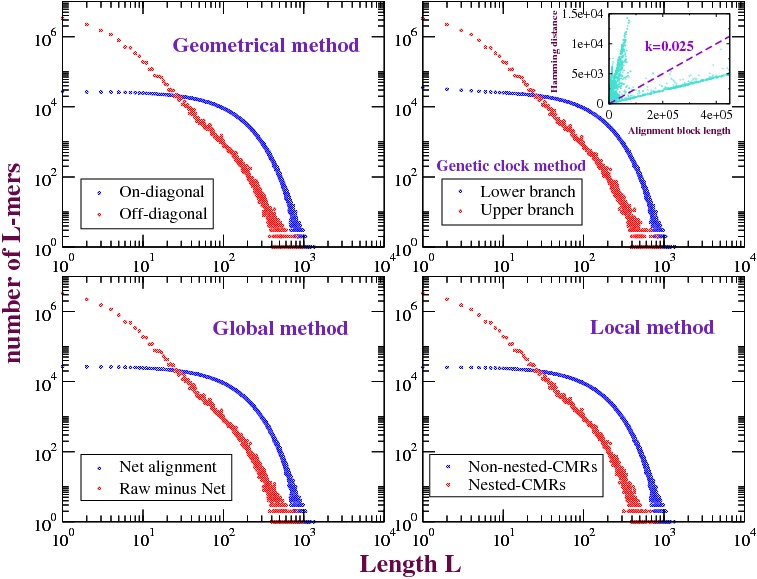}
\figcaption{Distributions of the ``orthologs" and ``paralogs" in our ``synthetic" alignment, separated by different methods.} 
\label{syndis}
\end{center}

\begin{table}[h]
\centering
\begin{tabular}{|c|c|C{0.2\textwidth}|C{0.2\textwidth}|c|}
\hline
\textbf{Methods}&\textbf{Subsets}&\textbf{numbers of orthologs}&\textbf{numbers of paralogs}&\textbf{Error (\%)}\\\hline
&On-diagonal&2456358&0&0\\\cline{2-5}
\raisebox{1.8ex}[0pt]{\textbf{``Geometrical"}}&Off-diagonal&0&12169905&0\\\hline
\textbf{``Genetic clock"}&Lower branch&2445738&62405&$2.49\%$\\\cline{2-5}
({\small ratio threshold: 0.025})&Upper branch&10642&12107501&$0.09\%$\\\hline
&Net alignment&2456357&17&$0.0007\%$\\\cline{2-5}
\raisebox{1.8ex}[0pt]{\textbf{``Global"}}&RMN alignment&7&12169893&$0.00006\%$\\\hline
&Non-nested-CMRs&2412292&11854&$0.489\%$\\\cline{2-5}
\raisebox{1.8ex}[0pt]{\textbf{``Local"}}&Nested-CMRs&44073&12158052&$0.361\%$\\\hline
\end{tabular}
\caption{Identification of ``orthologs" and ``paralogs" in the synthetic alignment by methods \textit{\ref{ms1}} -- \textit{\ref{ms4}}.}
\label{tb1}
\end{table}

Because evolution is simulated according to the RUPM, the orthologs and paralogs in this synthetic alignment can be identified solely by their locations within the aligned sequences and we can use this synthetic alignment to examine the reliability of the methods \textit{\ref{ms1}} -- \textit{\ref{ms4}} above. Figure \ref{syndis} illustrates the distributions of the ``orthologs" and ``paralogs" from our synthetic alignment, as separated by each of our four methods; evidently all of them are effective at separating the exponential from the power-law, as can also be seen from Table \ref{tb1}. Relative to the ``geometrical" method \textit{\ref{ms1}}, which is -- \textit{for the RUPM model} -- perfect, the other methods also perform well.

\subsection{Other orthologous chromosome pairs from human and chimpanzee}

The calculations above were performed on human chromosome $1$ and chimpanzee chromosome $1$. Figure \ref{otherchrs} exhibits distributions of exact matches from net and raw minus net alignments of all pairs of orthologous chromosomes from human and chimp. Exponential distributions characterise the net alignments, and algebraic most of the raw minus net. Some chromosome pairs show exponential tails in raw minus net, for example, chromosome $16$ and chromosome Y; it happens that these two chromosomes appear to contain more repetitive sequences than other chromosomes (data not shown); however, further understanding awaits future research.

\subsection{When species become more distantly related}

Heretofore we have addressed only the human-chimpanzee alignment. Whether our conclusions apply equally well to other genome pairs with similar evolutionary distances remains to be seen. Figure \ref{hpmc} shows the distributions of exact matches in alignments between human ($hg19$) chromosome $1$ and orthologous chromosomes selected from the Venter, chimpanzee, gorilla, orangutan, macaca and marmoset genomes. We choose for each species the orthologous cognate as the chromosome that shares the most orthologous genes with human chromosome $1$ according to Ensembl Biomart (data not shown). For a more distant genome, mouse chromosome $1$ is aligned to human chromosome $1$; it carries on the order of $1/4$ of orthologs between human chromosome $1$ and the mouse genome (see e.g. the human-mouse synteny map, http://cinteny.cchmc.org/doc/wholegenome.php). As the species pair diverges, distributions of shared sequences gradually cross over from exponential to algebraic. This crossover remains to be accounted for.

\begin{landscape}
\begin{center} 
\makebox[\textwidth][c]{\includegraphics[width=1.8\textwidth]{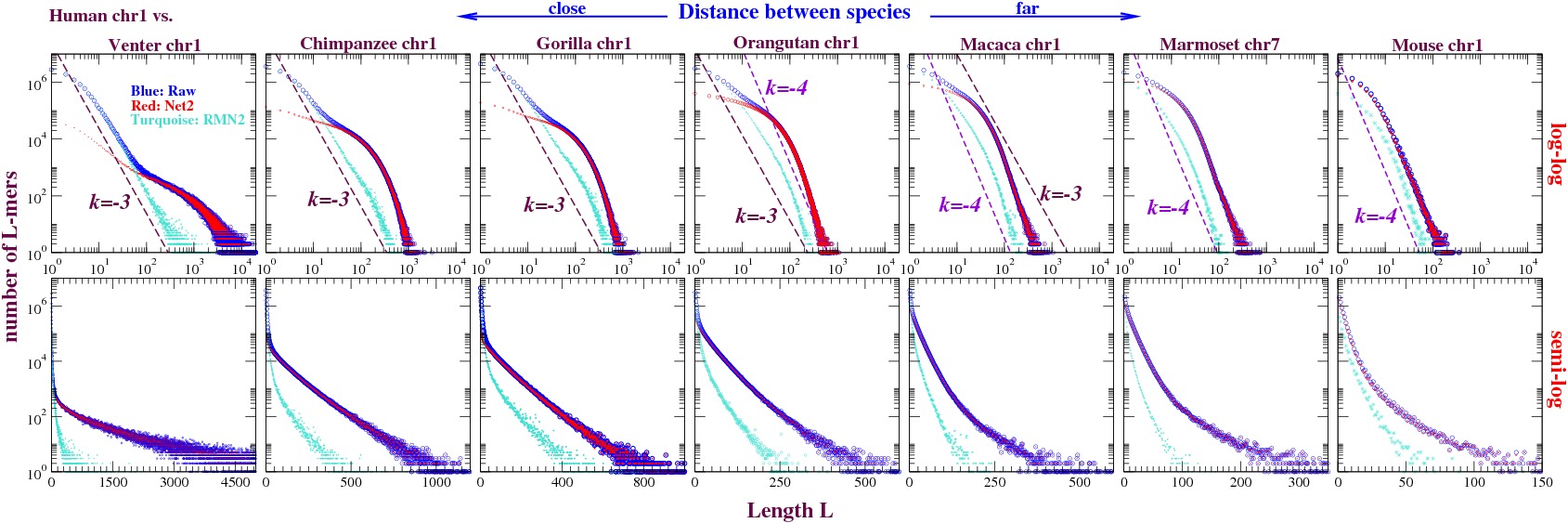}}
\figcaption{Distributions of exact matches from raw, net and raw minus net alignments of human chromosome $1$ versus the corresponding orthologous chromosomes of respectively Venter, chimpanzee, gorilla, orangutan, macaca, marmoset; and in the rightmost panel versus mouse chromosome $1$.}
\label{hpmc}
\end{center}
\end{landscape}

\section{Discussion} 

The quantitative study of monoscale substitution/duplication dynamics was revitalised by the work of H.C. Lee and collaborators with their apt characterisation of ``nature as the blind plagiariser" \cite{HCLee}. Although these authors did not investigate the steady state duplication length distributions yielded by their models, subsequent research revealed that similar classes of models yield algebraic length distributions that resemble those often exhibited by duplicated sequence in self-alignment and self-intersection of natural genomes \cite{GM,KM2011,stickbreak}. Algebraic distributions of conserved sequence lengths among distantly related genomes had been observed earlier.

This manuscript extends the characterisation of sequence length distributions to a pair of closely related genomes, those of human and chimp, where both conserved sequence lengths and duplicated sequence lengths can be simultaneously computed. In \textbf{Results} we demonstrated that the human chromosome $1$ -- chimpanzee chromosome $1$ alignment can be decomposed into two subsets, one with an exponential length distribution, the other an (approximately) algebraic length distribution. Our calculations also suggest that the algebraic length distribution is composed primarily of duplicated sequence including but not limited to paralogous genes, whereas the exponential length distribution is mainly composed of matches between orthologous chromosomal regions.

A neutral substitution model in the absence of selection is expected to yield an exponential length distribution for sequence conserved between two genomes. The phenomenon is quantitatively and conservatively thought of as a Bernoulli process; the exponential arises from the length distribution of head runs when flipping a biased coin \cite{AW}, and the exponential underlies most null models of sequence similarity in comparative genomics. It is not understood the extent to which an exponential is expected in (say) human/chimpanzee alignment, or whether -- since we are not chimpanzees -- an exponential is unexpected because of selection.

\subsection{Orthology and paralogy}
\label{oplg}

Chromosomal regions, within or across species, that have common ancestry are said to be \textit{homologs} \cite{homology}. Homologs can be further identified as \textit{orthologs} if they diverged via evolutionary speciation, or \textit{paralogs} if they diverged via sequence duplication \cite{AD,orthology}. Orthology and paralogy can in principle be defined for all sequences within a genome, but in practice most on-line databases consist only of protein-coding genes. Because of gene duplication and genome rearrangement, the ancestry of a given gene may be difficult to ascertain with high confidence, and ortholog/paralog classification can be ambiguous. Phylogenetic analysis of the gene lineage is customarily believed to enable the strongest discrimination between orthology and paralogy.

A standard approach to orthology and paralogy is to argue that within a given genome pair, orthologs tend to be those homologs that diverged least \cite{AD}. Duplication subsequent to speciation generates ``mother" and ``daughter" copies (known as ``in-paralogs") that exhibit congruent divergence from their cognate orthologs. This sequence of events yields so-called ``co-orthology" among in-paralogs \cite{Kristensen}. Co-orthology can be further refined to ``primary orthology" and ``secondary orthology" \cite{Han}. Our preliminary calculations suggest that in the human-chimpanzee alignment, \textit{primary} orthologs dominate the exponential length distribution, but \textit{secondary orthologs} merge with paralogs into the power-law length distribution.

\subsection{Approximate matching}
\label{apprmat}

In the plots above we study continuous (uninterrupted) matching runs of bases (CMRs), where continuous matching runs are by definition terminated at mismatches or indels; these are exact matches; however, CMRs may also be defined according to approximate matching criteria. The following criteria are listed in order of decreasing stringency: 

\begin{enumerate} 
\renewcommand{\labelenumi}{\MakeUppercase{\roman{enumi}} :} 
\item Exact matches: Each of the four nucleotides (A,T,G,C) matches itself only; a mismatch or indel terminates a run of matches; 
\item A=G, C=T: In addition to the exact matches, A and G, C and T are also matched pairs; an indel or any mismatch involving other than an A/G or T/C pair terminates the run;
\item Indel-terminated matches: aligned but gap/insert-free sequence is taken as matching; only an indel terminates the run; 
\item Alignment blocks: High similarity local alignments returned by LASTZ that are separated from one another by un-alignable sequence. They span exact matches, mismatches and indels. 
\end{enumerate}   
 
CMR distributions obtained with criteria \uppercase \expandafter {\romannumeral 1} through \uppercase \expandafter {\romannumeral 4} display sufficient qualitative similarity to one another that only exact match distributions are displayed in this manuscript. An example for human-chimpanzee alignment can be found in the supplement (see figure \ref{pan1} in \ref{diffstra}); for other genome pairs corresponding plots may be found in \cite{GM,koroteev2013,JMrep,SHM,TM2014}.

It was observed for distant \textit{inter}-genome comparisons in \cite{SHM} and \cite{JMrep} that criterion \uppercase \expandafter {\romannumeral 2} matches -- in contrast to all other inexact base substitution matching conditions -- displace the algebraic distribution of exact matches to numbers and lengths greater by an order-of-magnitude, with minimal impact on the shape of the curve. Were these C$\Rightarrow$T/G$\Rightarrow$A substitutions neutral, an exponential would have been anticipated. Yet, a qualitatively similar phenomenon (criterion \uppercase \expandafter {\romannumeral 2} shifts algebraic criterion \uppercase \expandafter {\romannumeral 1} curves to larger numbers and greater lengths, with minimal impact on shape) is observed for duplications \textit{within} a genome \cite{GM,KM2011,TM2014}.

\subsection{A conjecture on the crossover of orthologous sequence from exponential to algebraic}

The qualitative parallels between distributions of exact and inexact matches in duplicated sequence versus conserved sequence -- discussed in the previous section -- suggest to us that the mechanisms behind them share common features. Subsequent to  our original computations \cite{JMrep,SHM}, the portfolio of fully sequenced genomes has expanded vastly, and a variety of genome pairs exhibiting exact match length distributions with power laws close to $-3$ have emerged (jm, unpublished). This leads one of us (jm) to conjecture, supported by preliminary numerical calculations, a class of models that can account qualitatively for these observations.

The proposed class of models builds on the notion of sequence dynamics as fragmentation of a steady source \cite{koroteev2013,KRB,Kuhn,stickbreak,ZM}. The biological realisation of a mean steady source of newly duplicated sequence is readily plausible; its counterpart for sequence conservation may be more speculative. For sequence conservation, it is suggested that the counterpart of duplication is the steady generation of novel constrained sequences on which the constraints are newly relaxed. A few years ago, this notion seemed implausible, as the consensus among most biologists was that new functionalities arise through sequence duplication; however, recently evidence has emerged for alternative routes \cite{capra,ponting}. How much sequence arises through these alternative routes -- and what constitutes them -- is still unclear; for our purposes, it is not necessary to be too specific about details of any mechanism. Rather, we regard adaptation on the sequence level as a process of steady production (over evolutionary timescales) of novel sequence that serves novel functionalities, coupled with relaxation of or loss of constraint on sequences whose functionalities have become obsolete.

The latter yields a steady source of newly unconstrained sequence in the common ancestors that is reflected in descendants by randomly fragmented subsequences, as indicated in figure \ref{d9}. In figure \ref{d9} (a), the opaque coloured blocks represent newly duplicated sequence within a single lineage. The fading colour indicates the loss of homology between a duplicate sequence and its source as random local mutations fragment the matches. The time elapsed between the given duplication event and the present, governs the extent of fragmentation of the given duplicate. In figure \ref{d9} (b), the opaque coloured blocks represent sequence -- not necessarily duplicated -- on which selection has been newly lost. The faded colour indicates the loss of homology over time, as the newly unconstrained sequence accumulates random local mutations that fragment the matches. The evolutionary distance between a pair of genomes at the leaves of the tree -- reflecting the time elapsed between loss of constraint on the given sequence and the present -- governs the extent of fragmentation of the given sequence.

\begin{center} 
\makebox[\textwidth][c]{\includegraphics[height=0.6\textheight]{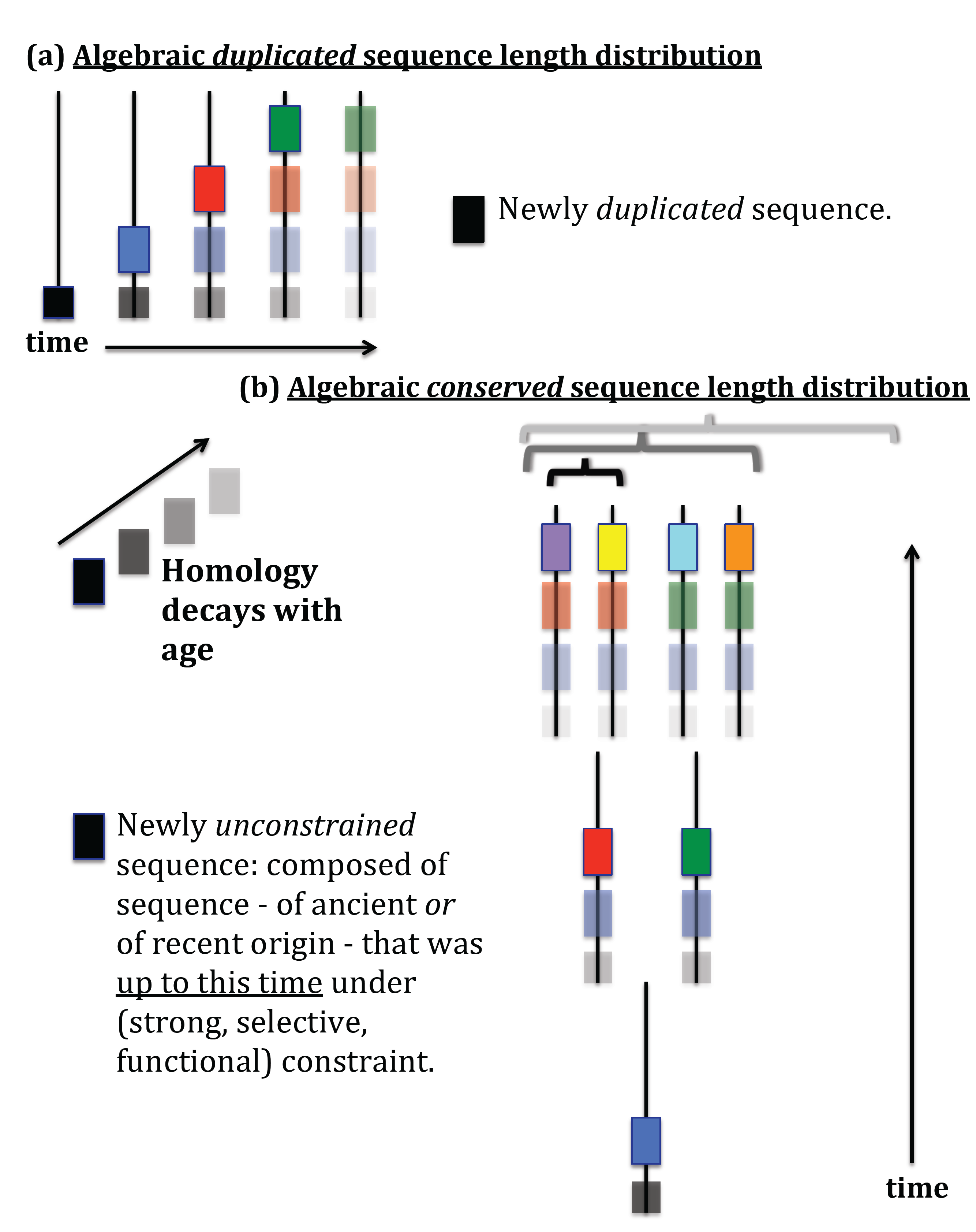}}
\figcaption{Schematic illustration of how a steady source of homologous sequence, subject to local mutation such as random base substitution, can lead to stationary algebraic distributions of homologous sequence length. (a) Solid coloured blocks represent newly duplicated sequence within a single lineage. The fading colour indicates the loss of homology between a duplicate sequence and its source as random local mutations fragment the matches. The time elapsed between the given duplication event and the present, governs the extent of fragmentation of the given duplicate. (b) Solid coloured blocks represent sequence -- not necessarily duplicated -- on which constraint has been newly lost. The faded colour indicates the loss of homology over time, as the newly unconstrained sequence accumulates random local mutations that fragment the matches. The time elapsed between the loss of constraint on the given sequence and the present, governs the extent of fragmentation of the given sequence. The braces indicate how this time is reflected in evolutionary distance: nearby leaves (black brace) are dominated by recent mutations of unconstrained sequence and yield an exponential distribution; intermediate distances (dark grey brace) are dominated by an algebraic distribution arising from successive losses of constraint; at still greater distances (light grey brace) the distribution exhibits increasingly steep tails as the overall amplitude attenuates into noise.}
\label{d9}
\end{center}

When comparing a \textit{pair} of present-day descendants of a common ancestor, fragmentation could be misinterpreted as representing the \textit{average} constraint on the sequence over all time; sequences that lost their constraints earlier appear subject to less constraint overall (are more fragmented) than sequences that lost their constraints more recently. Presumably, only suitable outgroup genomes can resolve this potential ambiguity. 

Observe that, in accord with figure \ref{d9} (b), recently diverged sequences (nearby branches) are expected to share exponentially distributed exact match lengths (because all the mutations breaking the matches occurred \textit{subsequent} to divergence); an intermediate regime to share algebraically distributed match lengths (the integral of fragment lengths arising from mutations that occurred \textit{before} divergence), in principle with power $-3$; as the branches separate further the amplitude of the distribution diminishes until matches are too sparse to infer its form.

In summary, the parallel between the algebraic distributions of duplicated and conserved sequence is that they both represent a signature of perpetual sequence \textit{turnover}; for conserved sequence, the \textit{turnover of functional sequence} in a continual process of expropriation, exploitation, and extinction. The latter conception is hardly novel, but the prospect of a quantitative measure of it (the exponent, presumably) could be illuminating.

\section{Conclusion}

Exponential length distributions between similar species and algebraic (power-law) length distributions between more distantly related species and within the alignment of a genome to itself have been previously observed. We have studied here the distribution of lengths of identical (and nearly identical) sequence shared between closely related organisms. A key contribution of our study is that the exponential distribution between closely related genomes turns out to be composed of two types of sequences: (1) orthologous sequences, which have an exponential distribution; (2) paralogous sequences, which have an algebraic (power-law) distribution.

Comparing human and chimpanzee, we explicitly distinguish orthologous from non-orthologous regions in a number of different ways, including known chromosome orthology; annotated orthologous regions in chromosomes; diagonal versus non-diagonal sectors of a dot plot; alignment similarity between human and chimpanzee; optimal chains of fragments aligned between orthologous chromosomes. For all such characterisations, we demonstrate exponentially distributed length segments for orthologous regions, and algebraic (power-law) distributed length segments for non-orthologous regions. Finally, we provide an \textit{in silico} demonstration of how such length distributions could have arisen through neutral evolution.

Recent models of neutral evolution proposed to explain algebraic distributions of duplicated sequence lengths often observed in natural genomes lead one to ask whether they can shed light on the evolution of duplications over evolutionary time scales \cite{koroteev2013,stickbreak}. Addressing this question suggests the investigation of duplicated sequences common to at least two different species. At the same time, observations from almost ten years ago of algebraic distributions of sequences conserved among multiple divergent species remain unaccounted for \cite{SHM}.

In this paper, we take some first steps in studying the evolution of the distribution of duplicated sequence lengths from self-alignment to alignment of two nearby species, human and chimpanzee. We describe a parameter-free method of extracting paralogs from LASTZ raw alignment of human and chimpanzee, based on nested and non-nested matches, that seems to reconstitute an approximately algebraic distribution of shared duplicate sequence lengths traceable to the self-alignment. Finally, we exhibit the evolution of orthologous sequence length distributions over a range of increasingly divergent species that spans the exponential and the algebraic, for which a mechanism is conjectured.

As observed in \cite{SHM} (twenty years after the rigorous mathematics of \cite{AW}) pure exponentials may not be so easy to come by in natural genome sequences. Once that has been recognized, the relevant question shifts to ``under what circumstances do exponentials actually occur, and why or why not?" And if not, what takes their place and what does it tell us about sequence evolution? We hope that the work presented here will eventually lead to further insights into these questions.

\section*{Acknowledgement}

The authors gratefully acknowledge generous support from the Okinawa Institute of Science and Technology Graduate University to jm.
\section*{References}
\bibliography{reference}
\bibliographystyle{plain}

\renewcommand\thefigure{\textit{S\arabic{figure}}}
\renewcommand\thetable{\textit{S\arabic{table}}}
\setcounter{figure}{0}
\setcounter{table}{0}

\section*{Supplementary figures}

\begin{center} 
\makebox[\textwidth][c]{\includegraphics[width=\textwidth]{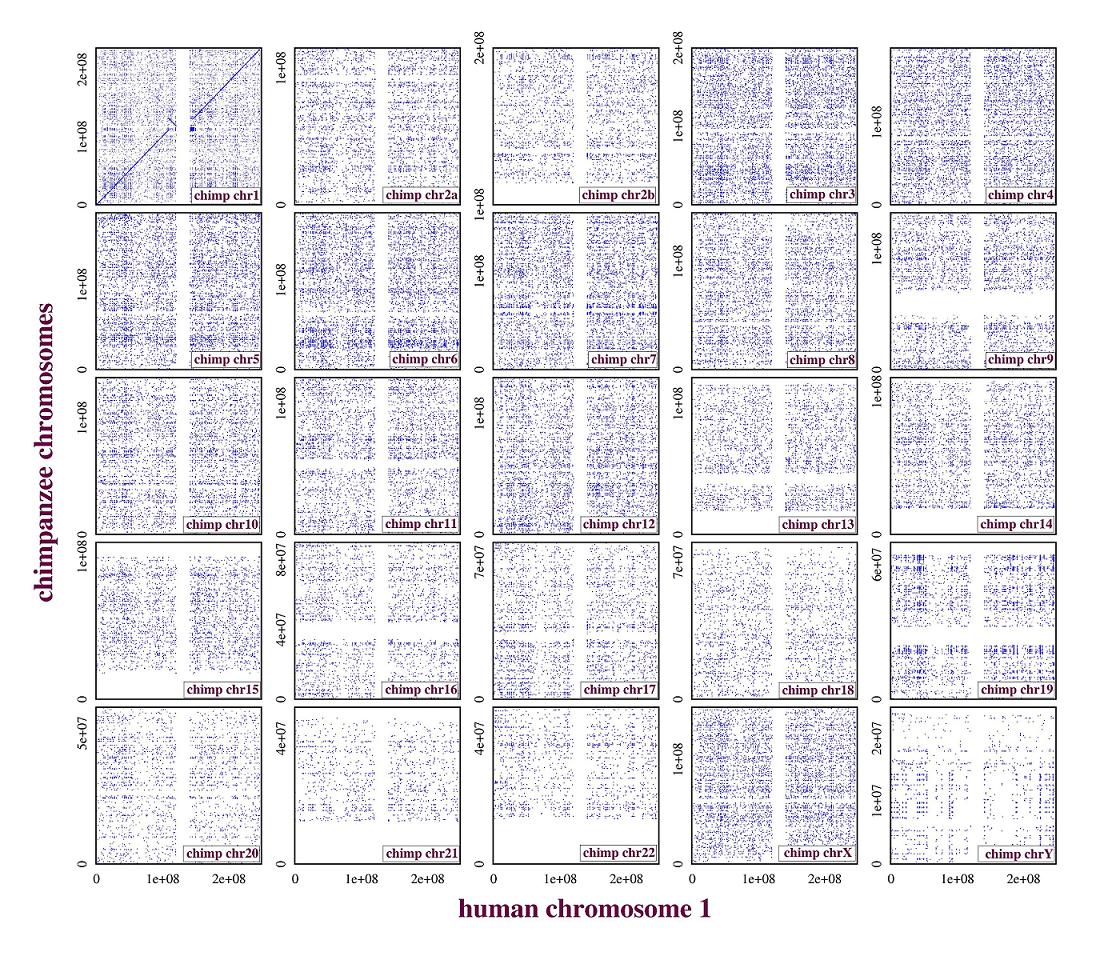}}
\figcaption{Dot plots for (soft repeat-masked LASTZ) raw alignments between human chromosome $1$ and each chimpanzee chromosome.}
\label{panalldp}
\end{center}

\begin{center} 
\makebox[\textwidth][c]{\includegraphics[width=0.75\textwidth]{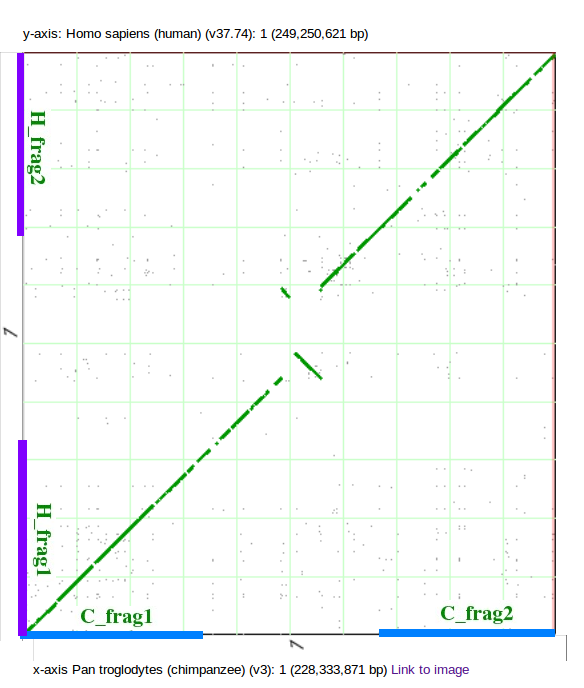}}
\figcaption{Syntenic dot plot for the CDS (protein-coding nucleotide sequences) between human chromosome $1$ and chimpanzee chromosome $1$, created by the SynMap tool in CoGe (\textit{http://genomevolution.org/CoGe/index.pl}). The horizontal blue and vertical violet bars indicate the locations of fragments $H\_frag1$, $H\_frag2$, $C\_frag1$ and $C\_frag2$ defined in figure \ref{hplocal}.}
\label{synmap}
\end{center}

\begin{center} 
\makebox[\textwidth][c]{\includegraphics[width=1.3\textwidth]{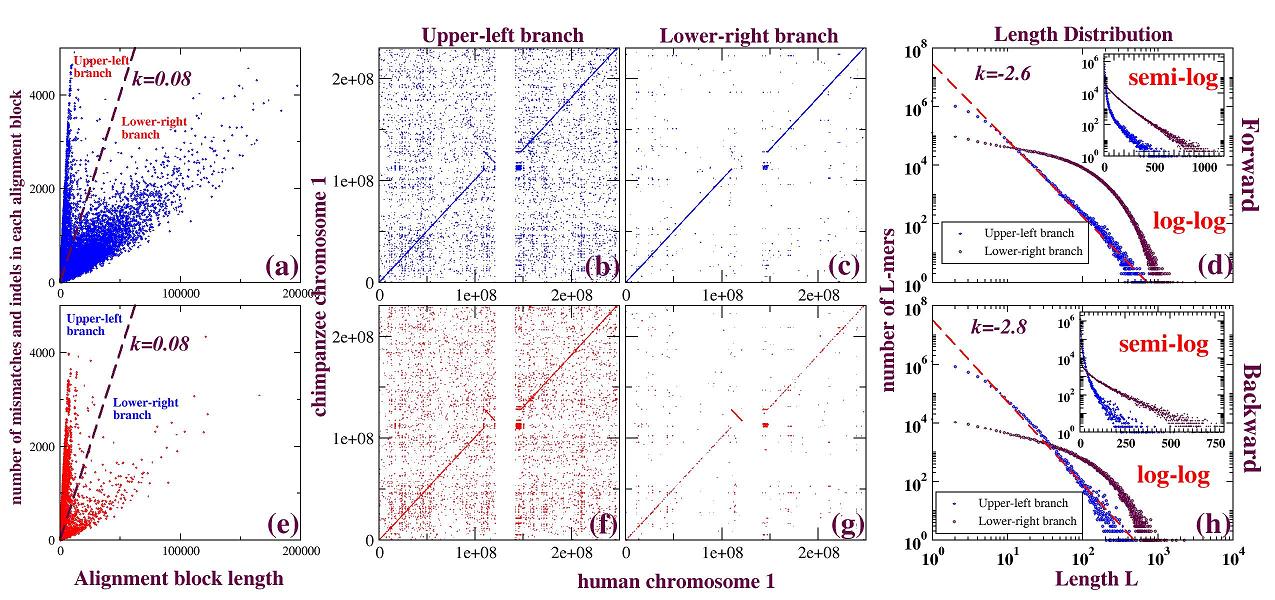}}
\figcaption{Same as figure \ref{DMR} in the main text, but displaying separately the forward and reverse alignments.}
\label{DMRFI}
\end{center}

\begin{center} 
\makebox[\textwidth][c]{\includegraphics[width=1.3\textwidth]{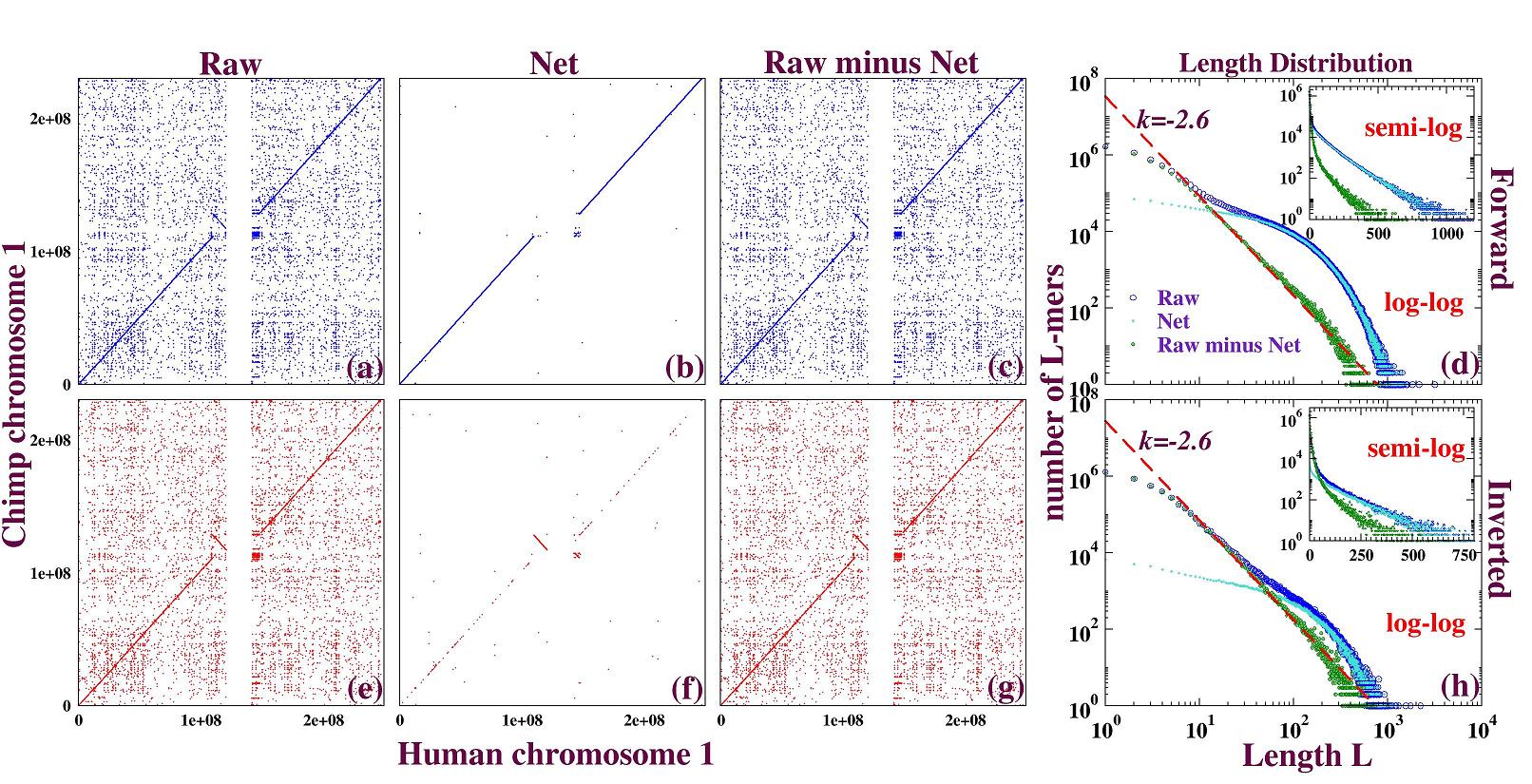}}
\figcaption{Same as figure \ref{RMN} in the main text, but displaying separately the forward and reverse alignments.}
\label{RMNFI}
\end{center}

\begin{center} 
\makebox[\textwidth][c]{\includegraphics[width=1.1\textwidth]{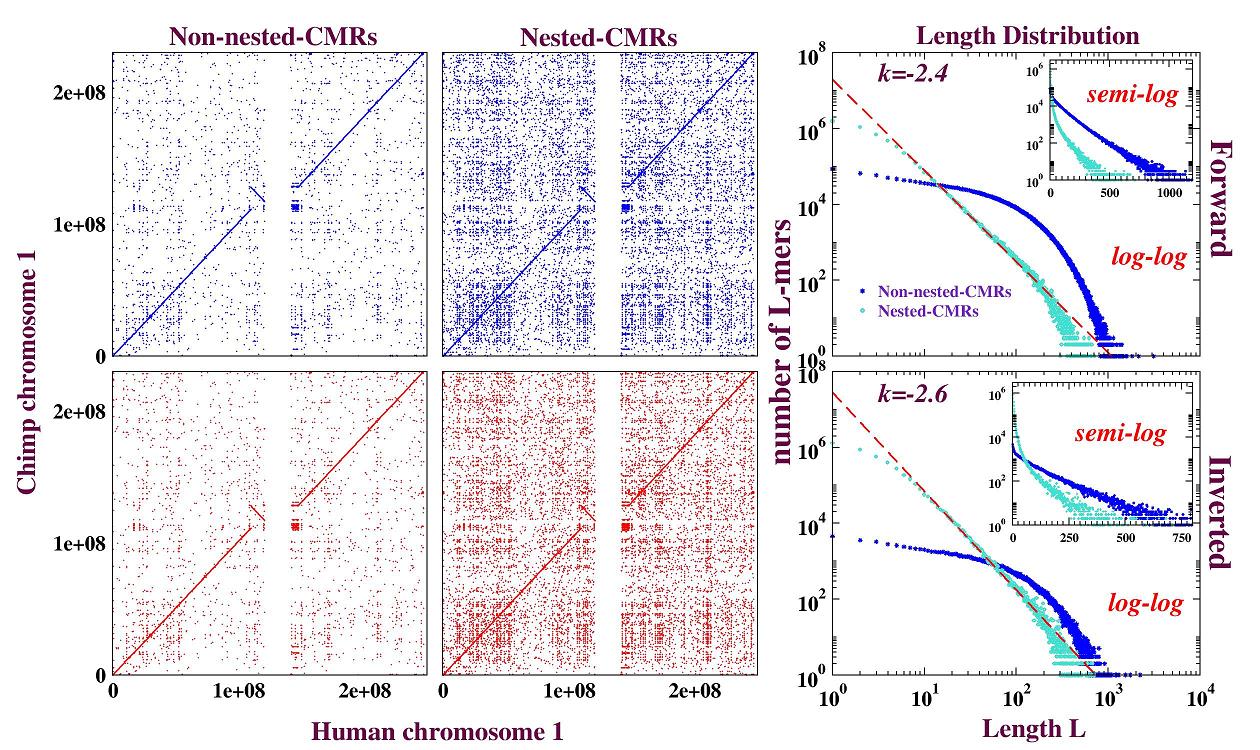}}   
\figcaption{Same as figure \ref{neov} in the main text, but displaying separately the forward and reverse alignments.}
\label{neovfi}
\end{center}

\begin{center} 
\makebox[\textwidth][c]{\includegraphics[width=1.1\textwidth]{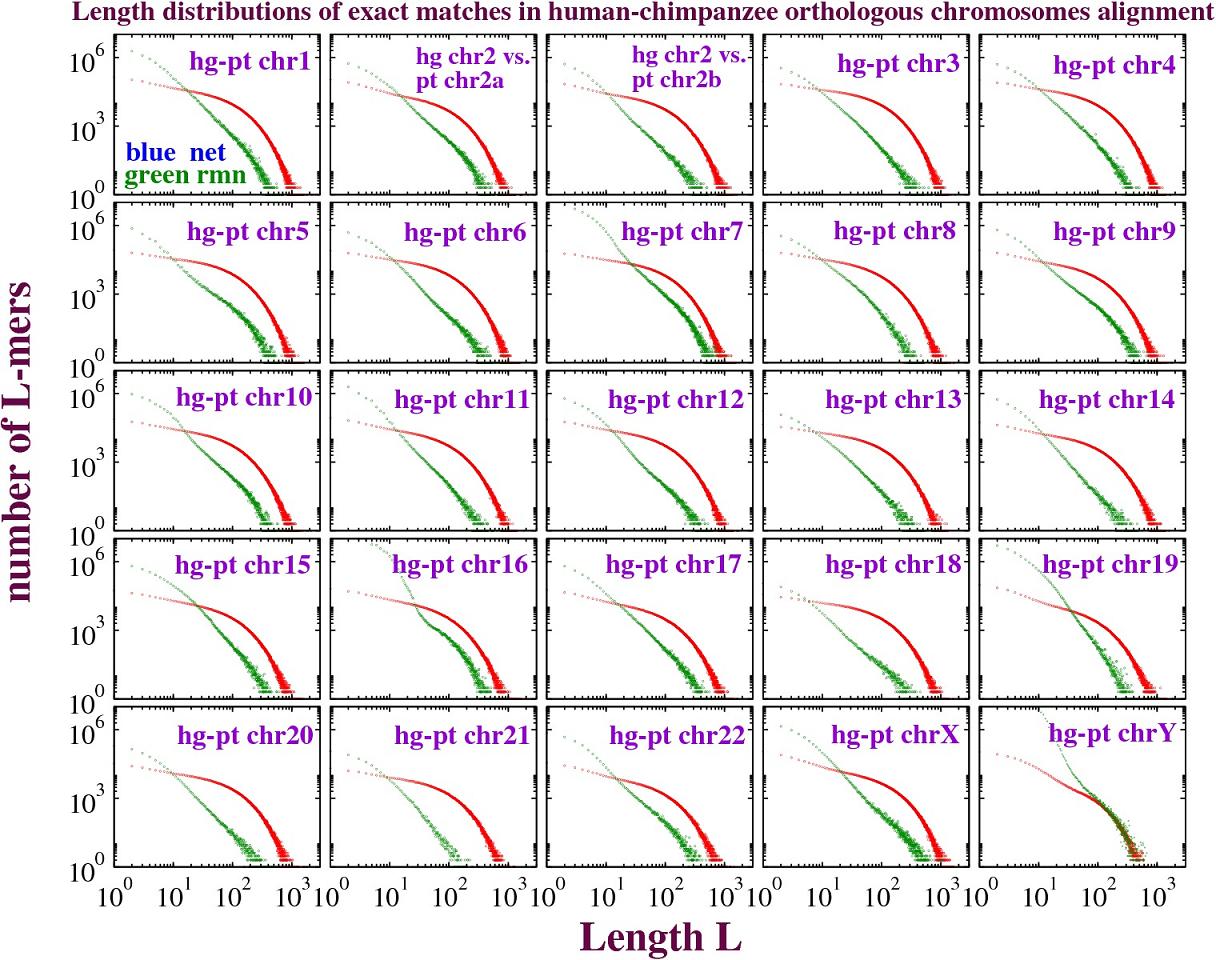}}   
\figcaption{(Length) distribution of exact matches in the net and RMN (raw minus net) alignments of chromosomes orthologous between human and chimpanzee.}
\label{otherchrs}
\end{center}

\section*{Supplementary Text}

\renewcommand\thesubsection{\textbf{Supplementary Text S\arabic{subsection}}}

\subsection{Control simulation for the synthetic alignment in section \textit{\ref{synsim}}}
\label{ctrlsim}

At the insistence of one of the referees, to confirm our interpretation of the simulation in section \textit{\ref{synsim}} we applied the numerical procedure described there to a random sequence of the same length as human chromosome $1$ and with lower-case letters at the same positions as they appear in soft-masked human chromosome $1$. Our simulation preserved the case of each letter, because unless this soft-masking was maintained, LASTZ was unable to complete an alignment of the descendent genomes. We applied $0.5\%$ substitutions per base independently to each of two copies of the random sequence and generated a LASTZ raw alignment between the mutated descendent genomes. ``Orthologs" and ``other matches" were identified via each of the methods \textit{\ref{ms1}} -- \textit{\ref{ms4}}. The distributions of exact matches are displayed in figure \ref{control}.

\begin{center} 
\includegraphics[width=0.7\textwidth]{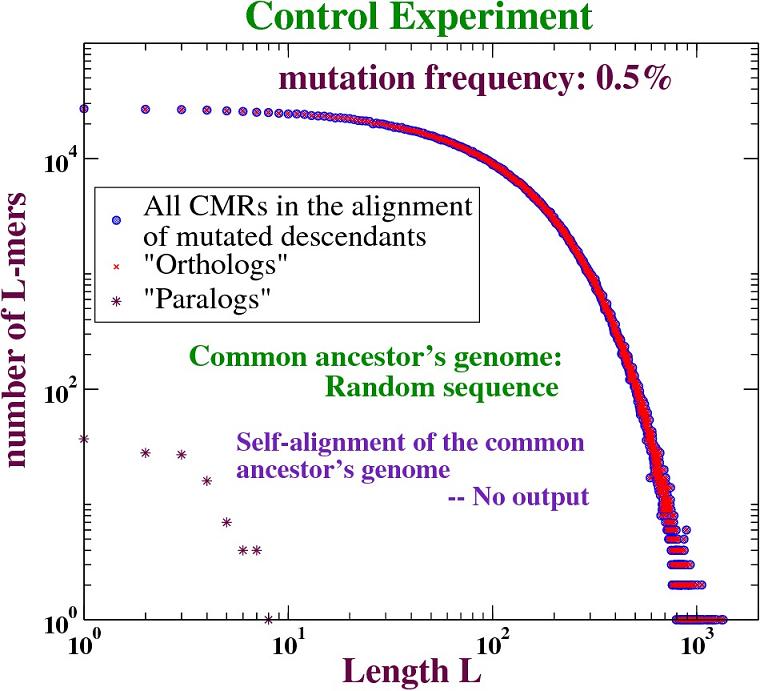}
\figcaption{A control simulation to figure \ref{synthe}. Random uncorrelated point mutations at a rate of $0.5\%$ per base are applied to produce two independent realisations of the RUPM on a randomly generated ``ancestral genome" of the same length as human chromosome $1$ and that are soft-masked at the same locations as in human chromosome $1$. Since the ancestral genome consists solely of independent uncorrelated random sequence, a power-law distribution of paralogs does not appear in this control simulation.}
\label{control}
\end{center}

Figure \ref{ctrldis} and table \ref{tb2} for the control simulation correspond respectively to figure \ref{syndis} and table \ref{tb1} of the text. ``Orthologs" in the synthetic and control simulations share qualitatively similar features, but because the random ancestral genome contains no segmental duplications, paralogs are absent in the control simulation (see inset in upper-right panel of figure \ref{ctrldis}, where the upper-left branch is missing). In figure \ref{control} any non-orthologous matches appear only by chance.

\begin{center} 
\includegraphics[width=0.7\textwidth]{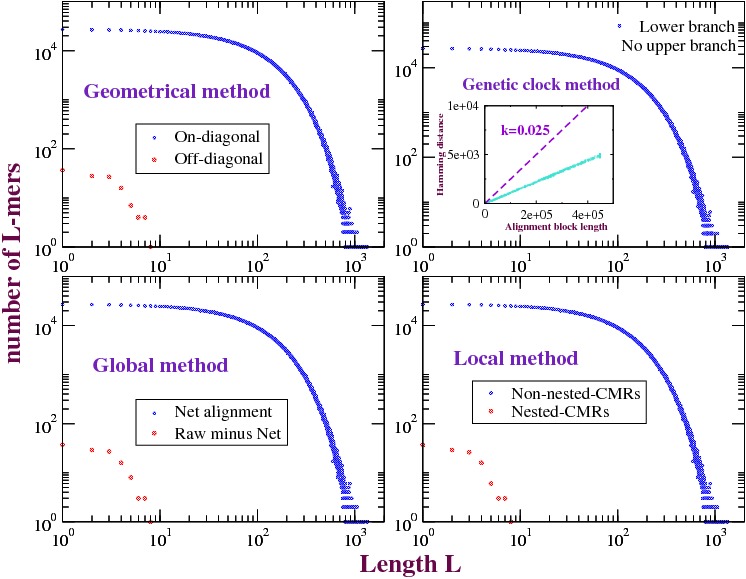}
\figcaption{Distributions of ``orthologs" and ``non-orthologous matches" in the control simulation as identified by each of the methods \textit{\ref{ms1}} -- \textit{\ref{ms4}}.} 
\label{ctrldis}
\end{center}

\begin{table}[h]
\centering
\begin{tabular}{|c|c|C{0.2\textwidth}|C{0.22\textwidth}|c|}
\hline
\textbf{Methods}&\textbf{Subsets}&\textbf{numbers of orthologs}&\textbf{numbers of non-orthologs}&\textbf{Error (\%)}\\\hline
&On-diagonal&2454994&0&0\\\cline{2-5}
\raisebox{1.8ex}[0pt]{\textbf{``Geometrical"}}&Off-diagonal&0&124&0\\\hline
\textbf{``Genetic clock"}&Lower branch&2455118&0&$0.005\%$\\\cline{2-5}
({\small ratio threshold: 0.025})&Upper branch&0&0&$0$\\\hline
&Net alignment&2454994&2&$0.00008\%$\\\cline{2-5}
\raisebox{1.8ex}[0pt]{\textbf{``Global"}}&RMN alignment&2&122&$1.613\%$\\\hline
&Non-nested-CMRs&2454993&4&$0.0002\%$\\\cline{2-5}
\raisebox{1.8ex}[0pt]{\textbf{``Local"}}&Nested-CMRs&1&120&$0.826\%$\\\hline
\end{tabular}
\caption{Identifications of orthologs and paralogs in the control simulation by methods \textit{\ref{ms1}} -- \textit{\ref{ms4}}.}
\label{tb2}
\end{table}

\subsection{Exponential distributions of CMRs counted by different matching criteria}
\label{diffstra}

Matching criteria \uppercase \expandafter {\romannumeral 1} through \uppercase \expandafter {\romannumeral 4} described in section \textit{\ref{apprmat}} successively relax the matching condition. CMRs counted according to a stricter criterion are contained within those counted according to a more relaxed criterion; therefore, locally and within an alignment block, CMRs counted according to different criteria exhibit a nested or hierarchical structure. Different matching criteria yield qualitatively similar length distributions, which suggests to us that the latter reflect some intrinsic features of the genomes rather than artefacts of matching criteria.

\begin{center} 
\includegraphics[width=\textwidth]{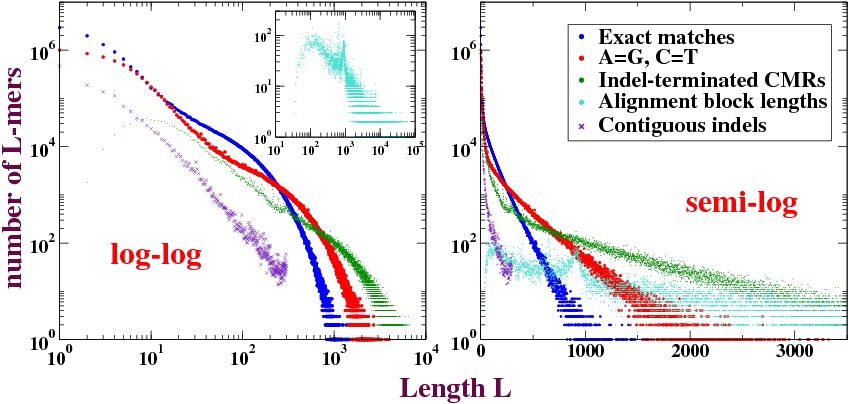}
\figcaption{Distributions of the contiguously matched runs counted by different matching criteria
in the (soft repeat-masked LASTZ) raw alignment between human chromosome $1$ and chimpanzee chromosome $1$.}
\label{pan1}
\end{center}

Figure \ref{pan1} shows the distributions of CMRs counted by different matching criteria from the human 
chromosome $1$ -- chimpanzee chromosome $1$ LASTZ raw alignment. Evidently, exact matches, A=G/C=T runs and 
indel-terminated runs all show exponential tails. As described in \cite{GM}, contiguous indels 
(successive insertions or deletions) yield algebraic length distributions.\\

The biological significance of matching criterion \uppercase \expandafter {\romannumeral 2}, transition 
(G$\Leftrightarrow$A, C$\Leftrightarrow$T), has been recognised since the discovery of the genetic code 
in the mid 20th century \cite{lehninger}. In the genetic code, it is evident that the 3rd base ``wobble" displays enhanced tolerance for transitions (they tend not to alter the amino acid encoded by a codon) over other kinds of substitutions. Furthermore, in RNA (DNA) secondary structure, the G:U (G:T) base-pair hydrogen bond plays a central role in stabilising duplex structures formed by all classes of RNAs \cite{varani}. Thus the tolerance in these contexts for G$\Leftrightarrow$A and C$\Leftrightarrow$T substitutions is reflected by functional selection.

On the other hand, the C$\Rightarrow$T mutation rate (and via subsequent mismatch repair, the G$\Rightarrow$A rate) is enhanced by an order of magnitude over other substitutions by the (selectively neutral) chemical process of deamination \cite{grauerli}. The effective pressure for C$\Rightarrow$T/G$\Rightarrow$A substitution is so strong that it is believed that certain specific mechanisms, such as A$\Rightarrow$G biased gene conversion, may have evolved to compensate for it \cite{fryxell}.

The biological relevance of criterion \uppercase \expandafter {\romannumeral 3} (indel terminated) is also widely recognised; consider, for example, the impact of an out-of-frame shift within protein-coding sequence. miRNAs are generally claimed to be insensitive to insertions or deletions within their interiors, but quite sensitive to insertions or deletions in their ``seed" regions; the latter observations have been applied quantitatively in \cite{lunter}.

\end{document}